\documentclass[prd,nofootinbib,twocolumn]{revtex4} 

\usepackage{natbib}
\usepackage{eufrak}
\usepackage{graphicx}
\usepackage{psfrag}

\usepackage[T1]{fontenc}

\def\O{{\cal O}}
\def\text#1{{\rm #1}}

\def\mb{\mbox}
\def\qaq{{\quad\text{and}\quad}}

\def\Lagr{{\Lambda}}
\def\E{{\cal E}}
\def\Om{\Omega}
\def\Lie{{\cal L}}
\def\Kepler{\mathrm{K}}

\def\K{{\cal K}}        

\def\X{{\!{\it X}}}     
\def\Y{{\!{\it Y}}}

\def\A{{\!{\it A}}}
\def\B{{\!{\it B}}}

\def\n{\mathrm{n}}         
\def\p{\mathrm{p}}         

\def\pn{p^\n}          
\def\pp{p^\p}         

\def\a{\alpha}   
\def\b{\beta}
\def\c{\gamma}

\def\C{{\cal C}}        

\def\eul{{\mathfrak{n}}}      

\def\enth{{\mu}}     
\def\entr{{\alpha}} 

\def\nix{{\rule{0em}{1em}}}

\def\cc{{c^2}}

\def\ph{\varphi}
\def\th{\theta}

\def\Mg{\mathcal{M}}   
\def\J{\mathcal{J}}

\def\kapn{\kappa_\n}
\def\kapp{\kappa_\p}
\def\kapnp{\kappa_{\n\p}}
\def\kapD{\kappa_{\!_\Delta}}

\def\eps{\varepsilon}
\def\xp{{x_\p}}

\def\stat{{(0)}}        
\def\snd{{(2)}}         
\def\fth{{(4)}}         

\def\S{\mathcal{S}}
\def\Et{\widetilde{E}}
\def\Eh{\hat{E}}
\def\D{\mathcal{D}}

\def\L{\mathrm{L}}
\def\sr{\mathrm{sr}}

\def\R{\mathcal{R}}  
\def\d{\delta}
\def\rel{\diamond}

\def\nuc{\mathrm{nucl}}
\def\fmq{\mathrm{fm}^{-3}}

\def\of{\mathrm{1f}}
\def\tf{\mathrm{2f}}

\def\pol{\mathrm{pol}}
\def\equ{\mathrm{eq}}

\def\Msol{M_\odot}
\def\FM{\mathrm{FM}}
\def\FCD{\mathrm{FCD}}

\def\nalpha{\bar{\alpha}}   

\begin{document}

\title{Relativistic numerical models for stationary superfluid Neutron Stars}

\author{Reinhard Prix}
\affiliation{Max-Planck-Institut f\"ur Gravitationsphysik,
Albert-Einstein-Institut, Am M\"uhlenberg 1,
D-14476 Golm, Germany}
\author{J\'er\^ome Novak}
\affiliation{Laboratoire de l'Univers et de ses Th\'eories,
  Observatoire de Paris, F-92195 Meudon Cedex, France}
\author{G.~L.~Comer}
\affiliation{Department of Physics, Saint Louis University, St.~Louis,
  MO, 63156-0907, USA}
\date{Jan 24, 2005}
        
\begin{abstract}
We have developed a theoretical model and a numerical code for stationary
rotating superfluid neutron stars in full general relativity.   
The underlying two-fluid model is based on Carter's covariant
multi-fluid hydrodynamic formalism.   
The two fluids, representing the superfluid neutrons on one hand, and the 
protons and electrons on the other, are restricted to uniform
rotation around a common axis, but are allowed to have different
rotation rates.    
We have performed extensive tests of the numerical code, including
quantitative comparisons to previous approximative results for these
models. The results presented here are the first ``exact''
calculations of such models in the sense that no approximations (other
than that inherent in a discretized numerical treatment) are used.
Using this code we reconfirm the existence of prolate-oblate shaped
configurations. We studied the dependency of the Kepler rotation limit
and of the mass-density relation on the relative rotation rate.
We further demonstrate how one can simulate a (albeit fluid)
neutron-star ``crust'' by letting one fluid extend further outwards
than the other, which results in interesting cases where the Kepler
limit is actually determined by the outermost but \emph{slower} fluid.
\end{abstract}

\maketitle

\section{Introduction}

The aim of this work is to calculate fully relativistic stationary
models of superfluid neutron stars including all non-dissipative
couplings between the two fluids induced by the equation of state
(EOS), in particular the entrainment effect. In addition to studying
the stationary properties 
of relativistic superfluid neutron stars, these models can serve as
the unperturbed initial state in a dynamical study of neutron star
oscillations, neutron star collapse to a black hole, or as a starting
point in studying pulsar glitch-models.  

Neutron stars are fascinating astrophysical objects: on one hand they
represent a formidable ``laboratory'' of fundamental physics, as
the composition and equation of state of their inner core still lies
beyond the reach of experimental and theoretical physics.
On the other hand, the advent of increasingly sensitive gravitational
wave detectors promises to open a new observational window
on neutron stars, which will allow us to gain new insights
into these still rather poorly understood objects. 
Gravitational wave astronomy could represent the first
opportunity to observe neutron star oscillations, providing a new
view on their inner dynamics. Considering the success of classical
terrestrial seismology and astero-seismology of the sun and of
main-sequence stars, one could expect this to result in substantial
progress in our understanding of the dynamics and composition of 
neutron stars.  

Additionally, observing quasi-permanent quadrupolar deformations
(``mountains'') on neutron stars via gravitational waves\footnote{Note
  that this search has already begun, see
  \cite{lsc:_settin_upper_limit} for a discussion and first results} 
will give valuable complementary information about their rotational
behavior, which is currently only observable via their electromagnetic
pulses.

Most theoretical studies of neutron star dynamics have relied on
rather simplistic single-fluid models. 
In this work we attempt a more realistic description of neutron stars
by taking their superfluidity into account via the use of a two-fluid
model. Neutrons and protons in neutron stars are predicted to 
be superfluid (e.g.~see~\cite{baldo92:_superfl_neutr_star_matter,sjoberg76:_effect_mass}), 
and this feature forms a fundamental ingredient in the current
(albeit rudimentary) understanding of the glitch phenomenon
observed in pulsars 
(e.g.~see~\cite{link00:_probing_NS_glitches,carter00:_centr_buoyancy,acp03:_twostream_prl}). 
Due to the superfluidity and therefore lack of viscosity of the
neutrons in the crust and in the outer core, they can flow freely
through the other components. The remaining constituents
(i.e. crust-nuclei, electrons, muons and protons) are assumed to be 
``locked'' together on short timescales by viscosity and the magnetic
field. Thereby they form another fluid, which in the
following will be referred to as ``protons'' for simplicity. These
assumptions characterize the so-called two-fluid model of neutron
stars. These two fluids are strongly coupled by the
strong nuclear force acting between protons and neutrons, and
therefore a hydrodynamic two-fluid framework incorporating these
couplings is required for their description. This framework will be
presented in the next section. Recently it was pointed out that such
a two-fluid system can be subject to a two-stream instability if the
relative velocity of the two fluids exceeds a critical velocity
\citep{acp04:_twostream}. This could therefore be relevant in neutron
stars and might be related to the glitch phenomenon\citep{acp03:_twostream_prl},
which provides another motivation for studying the properties of such
two-fluid systems. 

In this paper we study the stationary structure of such two-fluid
models, in which the two fluids are restricted to uniform rotation
around a common axis, but allowing for two different rotation rates.
This neutron star model was first studied quantitatively by
\citet{prix99:_slowl_rotat_two_fl_ns} in the Newtonian context
using a generalized Chandrasekhar-Milne slow-rotation approximation, 
and neglecting the direct interactions between the two fluids. 
\citet{andersson01:_slowl_rotat_GR_superfl_NS} used Hartle's
variant of the slow-rotation approximation to study this model in
general relativity. \citet{prix02:_slow_rot_ns_entrain} further
extended the Newtonian study to fully include all (non-dissipative)
couplings via entrainment and the nuclear ``symmetry-energy'', and
they found an analytic solution for a subclass of two-fluid equations
of state (which generalizes the $P\propto \rho^2$-type polytropes).
More recently, \citet{yoshida04:_rapid_newton} have devised an
alternative approach in the Newtonian case, by treating only the
relative rotation between the two fluids as small, while allowing
for fast rotation of the neutron star as a whole.
Furthermore, \citet{comer04:_slowl} has recently used the relativistic
slow-rotation approximation to study the properties of the first
available fully relativistic two-fluid EOS incorporating
entrainment, which was derived by \citet{comer03:_relat_mean_field}.

Here we present a generally relativistic numerical code for solving
the full two-fluid model without approximations.
A preliminary progress-report on the development of this code, and
some early results were presented in \citep{prix02:_relativistic_sf_ns}. 

While our model and code allow in principle for any given two-fluid
equation of state (EOS), for the sake of simplicity and a better
numerical convergence we restrict ourselves in this paper to the use
of a (rather general) class of two-fluid ``polytropes''. This choice is
also motivated by the lack of a useful two-fluid neutron star
equation of state in the literature, especially concerning the aspect
of entrainment.  Even though \citet{comer03:_relat_mean_field} have a 
fully relativistic model that includes entrainment, it has not yet 
been developed to the point that it will produce a tabular equation 
of state that could be used in our code.  
We expect the qualitative features of our model to be well
represented by the analytic EOS used in this work. 

The plan of this paper is as follows: 
In section~\ref{sec:canonical-two-fluid} we introduce the formalism
and notation of  covariant two-fluid hydrodynamics. 
In section~\ref{sec:stat-axisymm-conf} we discuss the specialization
to an axisymmetric and stationary system, and we introduce the $3+1$
framework for Einstein's equations. 
In section~\ref{sec:numerical-procedure} we describe the numerical 
procedure for solving the resulting elliptical system of equations.
The tests performed on the numerical code are discussed in
section~\ref{sec:tests-numerical-code}, and our numerical results are
presented in section~\ref{sec:numerical-results}. 
A discussion of this work is given in section~\ref{sec:summary}. 
In appendix~\ref{sec:newt-analyt-slow} we derive a new analytic
Newtonian slow-rotation solution, which is used for comparison to our
numerical results.  

\section{Canonical Two--Fluid Hydrodynamics}
\label{sec:canonical-two-fluid}

The general relativistic framework  for describing a coupled 
two-fluid system has been developed by Carter, Langlois and coworkers
\cite{carter89:_covar_theor_conduc,comer94:_hamil_superfluids,carter98:_relat_supercond_superfl,langlois98:_differ_rotat_superfl_ns},
based on an elegant variational principle. 
The same relativistic two-fluid model was used by
\citet{andersson01:_slowl_rotat_GR_superfl_NS} in their slow-rotation
description of superfluid neutron stars.

We consider a system consisting of two fluids, namely neutrons and
``protons'', which we label by $\n$ and $\p$ respectively. 
The kinematics of the two fluids is described by the two conserved
particle 4-currents $n_\n^\a$ and $n^\a_\p$, i.e.
\begin{equation}
\nabla_\a n_\n^\a = 0\,,\qaq 
\nabla_\a n_\p^\a = 0\,.
\label{equ1}
\end{equation}
The dynamics of the system is governed by a Lagrangian density
of the form $\Lagr(n^\a_\n, n^\a_\p)$.
Due to the requirement of covariance,  the scalar density $\Lagr$ can
only depend on  scalars, and we can form exactly three independent
scalar combinations out of $n^\a_n$ and $n^\a_\p$, for example
\begin{eqnarray}
n_\n^2 &\equiv& - {1\over \cc} g_{\a\b} n_\n^\a n_\n^\b\,,\nonumber\\
n_\p^2 &\equiv& - {1\over \cc} g_{\a\b} n_\p^\a n_\p^\b\,,\label{equScalars}\\
x^2    &\equiv& - {1\over \cc} g_{\a\b} n_\n^\a n_\p^\b\,,\nonumber
\end{eqnarray}
where $g_{\a\b}$ is the spacetime metric, so the Lagrangian density
can be written as
\begin{equation}
\Lagr(n^\a_\n, n^\a_\p) = - \E(n_\n^2, n_\p^2, x^2)\,,
\label{equ2}
\end{equation}
where $\E$ is a thermodynamic potential representing the
total energy density of the two--fluid system, or ``equation of state''.
Introducing the 4--velocities $u_\n^\a$, $u_\p^\a$ of the two fluids,
which satisfy the normalization conditions 
\begin{equation}
g_{\a\b}\, u_\n^\a \,u_\n^\b = -\cc\,,\qaq g_{\a\b}\, u_\p^\a \,u_\p^\b = -\cc\,,
\end{equation}
the particle 4--currents can be written as 
\begin{equation}
n_\n^\a = n_\n \,u_\n^\a \,,\qaq n_\p^\a = n_\p \,u_\p^\a\,,
\end{equation}
in terms of the neutron- and proton densities $n_\n$ and $n_\p$ respectively.
Variation of the Lagrangian density (\ref{equ2}) with respect to the
particle currents $n_\n^\a$ and $n_\p^\a$ defines the conjugate momenta
$\pn_\a$ and $\pp_\a$, namely
\begin{equation}
d\Lagr = \pn_\a \,d n_\n^\a + \pp_\a \,d n_\p^\a\,.
\label{eq:dLagr}
\end{equation}
Due to the covariance constraint (\ref{equ2}) we can further
express the conjugate momenta in terms of the currents as
\begin{eqnarray}
\pn_\a &=& \K^{\n\n} \,n_{\n\a} + \K^{\n\p} \,n_{\p\a}\,, \nonumber\\
\pp_\a &=& \K^{\p\n} \,n_{\n\a} + \K^{\p\p} \,n_{\p\a}\,, 
\label{equEntr}
\end{eqnarray}
where the symmetric ``entrainment matrix'' $\K^{\X\Y}$ is given by the
partial derivatives of \mb{$\E(n_\n^2, n_\p^2, x^2)$}, namely\footnote{The
  corresponding notation in   \citet{andersson01:_slowl_rotat_GR_superfl_NS} 
  is $n_\n^\a \rightarrow n^\a$, $n_\p^\a \rightarrow p^\a$,
  $u_\n^\a \rightarrow u^\a$, $u_\p^\a \rightarrow v^\a$, 
  $\K^{\n\p} \rightarrow {\cal A}  $, $\K^{\n\n} \rightarrow {\cal B}$, and 
  $\K^{\p\p} \rightarrow {\cal C}$.} 
\begin{equation}
\K^{\n\n} = {2\over \cc}{\partial \E \over \partial n_\n^2}\,,\quad
\K^{\p\p} = {2\over \cc}{\partial \E \over \partial n_\p^2}\,,\quad
\K^{\n\p} = {1\over\cc}{\partial \E \over \partial x^2}\,.
\label{equEntrMatrix}
\end{equation}
The equations of motion for the two fluids can be obtained from the
variational principle developed by \citet{carter83:_in_random_walk}.  
In the absence of direct dissipative forces acting between the two fluids
(e.g.~see~\citet{langlois98:_differ_rotat_superfl_ns}), the equations
of motion will then be found as\footnote{the square
  brackets denote averaged index anti-symmetrization, i.e. 
$2 \,v_{[a, b]} = v_{a b} - v_{b a}$. }
\begin{equation}
n_\n^\a \nabla^\nix_{[\a} \pn_{\b]} = 0\,,\qaq
n_\p^\a \nabla^\nix_{[\a} \pp_{\b]} = 0\,.
\label{equEOM}
\end{equation}
The energy--momentum tensor $T^{\a\b}$, which is derived from the
variational principle too, has the form 
\begin{equation}
{T^\a}_\b = n_\n^\a \,\pn_\b + n_\p^\a \,\pp_\b + \Psi g^\a_\b\,.
\label{equTmunu}
\end{equation}
If the equations of motion (\ref{equ1}) and (\ref{equEOM}) are satisfied, the
stress-energy tensor automatically satisfies \mb{$\nabla_\a   T^{\a\b}=0$},  
which is a Noether-type identity of the variational principle. The
generalised pressure $\Psi$ of the two--fluid system is 
defined by the thermodynamical identity 
\begin{equation}
\E + \Psi = -n_\n^\a \,\pn_\a - n_\p^\a\, \pp_\a \,,
\label{equTD}
\end{equation}
so $\Psi$ can be considered the Legendre-transform of the Lagrangian
density $\Lagr$. Using the entrainment relation (\ref{equEntr}), we
can rewrite this as 
\begin{equation}
{\E +\Psi \over \cc} =  \K^{\n\n} \,n_\n^2 + 2 \K^{\n\p} \,x^2 + \K^{\p\p} \,n_\p^2\,.
\label{equTD2}
\end{equation}
Instead of $x^2$ defined  in~(\ref{equScalars}), we will use a
physically more intuitive quantity as the third independent scalar,
namely the ``relative speed'' $\Delta$.  
We define the relative speed $\Delta$ as the norm of the neutron
velocity $u_\n^\a$ as seen in the frame of the protons $u_\p^\a$, or
vice versa. The corresponding relative Lorentz factor $\Gamma_\Delta$
is therefore given by
\begin{equation}
\Gamma_\Delta = - {1\over \cc}u_\n^\a u_\p^\b \,g_{\a\b} = {x^2 \over n_\n n_\p} =
\left(1 - {\Delta^2\over \cc} \right)^{-1/2}\,,
\end{equation}
and the relative speed $\Delta$ is expressible in terms of $x$ as
\begin{equation}
\Delta^2 = \cc \left[ 1 - \left( n_\n n_\p \over x^2 \right)^2 \right]\,.
\label{eq:DefDelta}
\end{equation}
In the case of co-moving fluids (i.e. $u_\n^\a = u_\p^\a$), we see
from~(\ref{equScalars}) that $x^2 = n_\n\,n_\p$, and so $\Delta=0$ as
expected. 

We can now equivalently consider the equation of state $\E$ as a
function of the form $\E(n_\n, n_\p, \Delta^2)$, for which the  first
law of thermodynamics reads as
\begin{equation}
d\E = \enth^\n dn_\n + \enth^\p dn_\p + \entr \,d\Delta^2\,, 
\label{equFirstLawII}
\end{equation}
closely analogous to the Newtonian formulation \cite{prix02:_slow_rot_ns_entrain}. 
The conjugate quantities defined in this equation are the entrainment
$\entr$ and the neutron- and proton chemical potentials $\enth^\n$ and $\enth^\p$
(sometimes also referred to as specific enthalpies, which is
equivalent in the zero-temperature case). 
It is often useful to characterize the entrainment by the
dimensionless entrainment numbers $\eps_\X$, which we define as
\begin{equation}
\eps_\X \equiv {2 \entr \over m^\X n_\X}\,,
\label{eq:10}
\end{equation}
where $m^\X$ is the particle rest-mass of the respective fluid, and
the fluid-index is $\X = \n, \p$ (no summation 
over $\X$).

The conjugate variables of (\ref{equFirstLawII}) can be expressed in
terms of the kinematic scalars and the entrainment matrix $\K^{\X\Y}$ as 
\begin{eqnarray}
\enth^\n &=& {\cc\over n_\n}\left(\K^{\n\n} n_\n^2 + \K^{\n\p} x^2\right)
= -u_\n^\a \pn_\a\,, \nonumber\\
\enth^\p &=& {\cc\over n_\p}\left(\K^{\p\p} n_\p^2 + \K^{\n\p} x^2\right)
= -u_\p^\a \pp_\a\,, \label{equEnthalpies}\\
\entr &=& {1\over 2} \K^{\n\p} n_\n n_\p \Gamma_\Delta^3\,.\nonumber
\end{eqnarray}
Using (\ref{eq:DefDelta}), the inverse relations can be obtained as 
\begin{eqnarray}
\K^{\n\n} &=& {\enth^\n \over n_\n \cc} - {2 \entr \over n_\n^2
\Gamma_\Delta^2} \nonumber\\
\K^{\p\p} &=& {\enth^\p \over n_\p \cc} - {2 \entr \over n_\p^2
\Gamma_\Delta^2} \label{equEntrMatrix2a}\,\\
\K^{\n\p}  &=& {2 \entr \over n_\n n_\p \Gamma_\Delta^3}\,,\nonumber
\end{eqnarray}
which reduces exactly to the corresponding relations in the Newtonian
limit \cite{prix02:_slow_rot_ns_entrain}, where $\Gamma_\Delta\rightarrow1$ and 
$\enth^\X \rightarrow m^\X \cc + \widehat{\mu}^\X$. 
In terms of these quantities, the generalised pressure $\Psi$
(\ref{equTD}) can also be written as 
\begin{equation}
\Psi = -\E  + n_\n \,\enth^\n + n_\p \,\enth^\p\,,
\label{equPressure2}
\end{equation}
which is the generalization of the thermodynamic Gibbs-Duhem relation
to a two-fluid system. 

\section{Stationary axisymmetric configurations}
\label{sec:stat-axisymm-conf}

\subsection{The metric}

Here and in the following we choose units such that $G = c = 1$ for
simplicity.  
We consider spacetimes that are stationary, axisymmetric, and
asymptotically flat. The symmetries of stationarity and axisymmetry
are associated with the existence of two Killing vector fields, 
one timelike at spatial infinity, $t^\a$, and one spacelike everywhere
and with closed orbits, $\varphi^\a$. 

It was shown by \citet{carter70:_commut} that under these
assumptions the Killing vectors commute, and one can choose an
adapted coordinate system $(t, x^1, x^2, \varphi)$, such that
\mb{$t^\a \partial_\a = \partial/\partial t$} and  \mb{$\varphi^\a
  \partial_a = \partial / \partial \varphi$}, i.e. 
\begin{equation}
t^\a = (1, 0, 0, 0)\,,\qaq \varphi^\a = (0, 0, 0, 1)\,.
\end{equation}
We choose the remaining two coordinates to be of spherical type,
i.e. $x^1=r$, $x^2=\theta$, and following
\citet{gourgoulhon99:_fast_rotation_strange_stars}, we fix the gauge
to be of maximal-slicing quasi-isotropic type (MSQI), for which the
line element reads as
\begin{eqnarray}
ds^2 &=& g_{\a\b}\,dx^\a dx^\b = -(N^2 - N_\varphi N^\varphi) \, dt^2 
- 2 N_\varphi \,d\varphi \,dt  \nonumber\\
& &+ A^2\left( dr^2 + r^2 d\theta^2 \right) + B^2 r^2
\sin^2\!\theta \,d\varphi^2\,,
\label{eq:7}
\end{eqnarray}
where the functions $N$, $N^\varphi$, $A$ and $B$ depend on $r$ and
$\theta$ only, and \mb{$N_\varphi \equiv g_{\varphi\varphi} N^\varphi$}.

\subsection{Fluid dynamics}
We assume the flow of the two fluids to be purely {\em axial} (i.e. no
convective meridional currents), so we can write the unit
4--velocities of the two fluids as
\begin{equation}
u_\n^\a = u_\n^t \,\zeta_\n^\a\,\qaq
u_\p^\a = u_\p^t \,\zeta_\p^\a\,,
\label{equ4Velocities}
\end{equation}
where the helical vectors $\zeta_\n^\a$ and $\zeta_\p^\a$ are
expressible in terms of the Killing vectors $t^\a$ and $\varphi^\a$ as 
\begin{equation}
\zeta_\n^\a = t^\a + \Om_\n \varphi^\a\,,\qaq
\zeta_\p^\a = t^\a + \Om_\p \varphi^\a\,,
\label{equXi}
\end{equation}
where the two rotation rates $\Om_\n$ and $\Om_\p$ are scalar functions,
which can only depend on $r$ and $\theta$.

Using Cartan's formula for the Lie derivative of a 1--form $p_\b$ with
respect to a vector--field $\xi^\a$, namely
\begin{equation}
\Lie_\xi \;p_\a = 2 \xi^\b \nabla_{[\b} p_{\a]} 
+ \nabla_\a\left( \xi^\b p_\b \right)\,,
\label{equCartan}
\end{equation}
we can rewrite the equations of motion (\ref{equEOM}) as
\begin{equation}
\Lie_{\zeta_\X} \p^\X_\a - \nabla_\a\left( \zeta^\b_\X \p^\X_\b \right) = 0\,.
\end{equation}
Linearity of the Lie
derivative together with (\ref{equXi}) and (\ref{equCartan}) allows us
to rewrite this as 
\begin{equation}
\Lie_t\, p^\X_\a + \Om_\X \Lie_\varphi\, p^\X_\a + 
\varphi^\b p^\X_\b\, \nabla_\a \Om_\X - \nabla_\a\left(\zeta_\X^\b
p^\X_\b\right) = 0\,.
\end{equation}
Stationarity and axisymmetry imply that the first two terms vanish,
and so the equations of motion for neutrons and protons are reduced to
\begin{eqnarray}
\pn_\varphi \nabla_\a \Om_\n = \nabla_\a\left(\zeta^\b_\n\, \pn_\b \right)\,,\quad
\pp_\varphi \nabla_\a \Om_\p = \nabla_\a\left(\zeta^\b_\p\, \pp_\b \right)\,.
\end{eqnarray}
In the general case of differential rotation, the integrability condition
of these equations are therefore 
\begin{equation}
\pn_\varphi = \pn_\varphi ( \Om_\n )\,,\qaq
\pp_\varphi = \pp_\varphi ( \Om_\p )\,,
\end{equation}
and the first integrals of motion are obtained as
\begin{eqnarray}
\pn_t + \Om_\n \pn_\varphi - \int^{\Om_\n} \pn_\varphi(\Om') d\Om'
&=& \text{const}^\n \,, \\
\pp_t + \Om_\p \pp_\varphi - \int^{\Om_\p} \pp_\varphi(\Om') d\Om'
&=& \text{const}^\p \,.
\end{eqnarray}
In the special case of {\em uniform rotation}, i.e. 
$\nabla_\a \Om_\X=0$, these first integrals reduce to
\begin{equation}
\pn_t + \Om_\n \pn_\varphi = \text{const}^\n\,,\qaq
\pp_t + \Om_\p \pp_\varphi =  \text{const}^\p \,,
\label{equFirstIntegral}
\end{equation}
which are equivalent to the expressions obtained by 
\citet{andersson01:_slowl_rotat_GR_superfl_NS}. We can further express
these first integrals in terms of the chemical potentials $\enth^\n$,
$\enth^\p$ of (\ref{equEnthalpies}), namely 
\begin{eqnarray}
\pn_t + \Om_\n \pn_\varphi &=& \zeta_\n^\a \pn_\a 
= - {1\over u_\n^t} \enth^\n = \textrm{const}^\n\,,
\label{equFirstIntegral2a}\\
\pp_t + \Om_\p \pp_\varphi &=& \zeta_\p^\a \pp_\a 
= - {1\over u_\p^t} \enth^\p = \textrm{const}^\p\,.
\label{equFirstIntegral2b}
\end{eqnarray}

\subsection{The $3+1$ decomposition}
We introduce the vector $\eul^\a$ as the unit normal to the spacelike
hypersurfaces $\Sigma_t$ defined by $t={\rm const}.$, namely
\begin{equation}
\eul_\a \equiv -N \nabla_\a t\,,
\end{equation} 
which defines the so--called {\em Eulerian observers} $\O_0$
following \citet{smarr_york78:_kinem}.
The induced metric $h_{\a\b}$ on the spacelike hypersurfaces $\Sigma_t$ is
given by the projection
\begin{equation}
h_{\a\b} \equiv g_{\a\b} + \eul_\a \eul_\b\,.
\end{equation}
The corresponding $3+1$ decomposition of the energy--momentum tensor
$T^{\a\b}$ reads as\footnote{Round brackets denote averaged symmetrization,
  i.e. $2 \,v_{(a, b)} = v_{a b} + v_{b a}$.}
\begin{equation}
T^{\a\b} = S^{\a\b} + 2 \eul^{(\a}J^{\b)} + E \eul^\a \eul^\b\,,
\label{equ3+1}
\end{equation}
where
\begin{equation}
E = \eul^\a T_{\a\b} \eul^\b\,,\quad
J_\a = - h^\c_\a T_{\c\b} \eul^\b\,,\quad
S_{\a\b} = h^\c_\a T_{\c\nu} h^\nu_\b\,,
\end{equation}
which can be interpreted as the energy, momentum and stress tensor as
measured by the Eulerian observers.
In the MSQI gauge~(\ref{eq:7}), we can explicitly express these
quantities as
\begin{equation}
E = N^2 T^{tt}\,,\quad J_i = N T^t_i \,,\quad
S^i_j = T^i_j - N^i T^t_j\,,
\end{equation}
where Latin indices $i,\,j$ denote the space-components $1,2,3$.
The Einstein equations in this formulation result in a set of four
elliptic equations for the metric potentials (see \cite{bonazzola93:_axisy}
and \cite{gourgoulhon99:_fast_rotation_strange_stars} for details), namely
\begin{eqnarray}
\Delta_3 \nu = 4\pi A^2 (E + S_i^i)  + A^2  K_{i j} K^{i j}
- \partial \nu \,\partial (\nu + \beta),\label{eq:Poisson3D}\\ 
\tilde{\Delta}_3 \widetilde{N}^\ph = -16\pi N A^2 \widetilde{J}^\ph 
- r \sin\th\, \partial N^\ph \partial (3\beta - \nu)\,, \label{eq:PoissonVect}\\
\Delta_2[(NB-1) r\sin\th]  = 8\pi N A^2 B r \sin\th \left( S_r^r + S_\th^\th \right),\\
\Delta_2 \left( \nu + \nalpha\right) = 
8\pi A^2 S_\ph^\ph + {3\over2} A^2 K_{i j}K^{i j} - \left(\partial \nu\right)^2\,,\label{eq:Poisson2D}
\end{eqnarray}
where we defined $\widetilde{N}^\ph \equiv r\sin\th\,N^\ph$ and
$\widetilde{J}^\ph \equiv r\sin\th\, J^\ph$. $\Delta_3$ and $\Delta_2$
are the flat three- and two-dimensional Laplace operators, whereas
$\tilde{\Delta}_3 = \Delta_3 - (r^2 \sin^2 \th)^{-1}$. We further
used the notation
\begin{equation}
\nu \equiv \ln N\,,\quad \nalpha \equiv \ln A \,,\quad \beta \equiv \ln B \,,
\end{equation}
and we define $\partial \nalpha\partial \beta$ as the flat-space scalar
product of two gradients, i.e.
\begin{equation}
\partial \nalpha \partial \beta \equiv \partial_r\nalpha \partial_r \beta +
{1\over r^2} \partial_\th \nalpha \partial_\th \beta\,.
\end{equation}
The only non-zero components of the extrinsic curvature $K_{i j}$ in our
spherical coordinate basis are given by 
\begin{equation}
K_{r\ph} = - {g_{\ph \ph} \over 2 N} \, \partial_r N^\ph\,,\;\;\textrm{and}\;\;
K_{\th\ph} = - {g_{\ph \ph} \over 2 N}\, \partial_\th N^\ph\,.
\end{equation}
We note that the gravitational mass $\Mg$, which is defined as the
(negative) coefficient of the term $1/r$ in an asymptotic expansion of
the ``gravitational potential'' $\log\,N$, can be expressed 
explicitly (see \cite{bonazzola93:_axisy}) as
\begin{equation}
\Mg = \int A^2 B \left[ N (E + S_i^i) + 2 B^2 \widetilde{N}^\ph
  \widetilde{J}^\ph \right]\, r^2 \sin\th \, dr\,d\th\,d\ph\,.
\label{eq:6}
\end{equation}
Here and in the following we will use $\Mg$ to denote the
gravitational mass, while $M$ will stand for the baryon mass.
The total angular momentum $\J$ is given by
\begin{equation}
\J = \int \left[ A^2 B^3 \, r\sin\th \widetilde{J}^\ph\right]\, r^2\sin\th\,dr\,d\th\,d\ph\,.
\end{equation}
The 2D- and 3D virial identities, which have been derived by Bonazzola
and Gourgoulhon \cite{gourgoulhon94:virial_grv3,bonazzola94:virial_grv2}, 
can serve as a useful check of consistency and precision of the
numerical results. The 2D virial identity (referred to as GRV2), which
derives from the Poisson-equation (\ref{eq:Poisson2D}), has the form 
\begin{equation}
\int \left[ 8\pi A^2 S_\ph^\ph + {3\over2} A^2 K_{i j} K^{i j} -
  (\partial \nu)^2 \right] \, r\,dr\,d\th = 0\,,
\label{eq:GRV2}
\end{equation}
while the 3D virial identity (GRV3), which reduces to the usual virial
theorem in the Newtonian limit, can be written as
\begin{eqnarray}
\int 4\pi A^2 B S_i^i \,d V + \int B\left[ {3\over4}\,A^2 K_{i j}K^{i j} 
 - (\partial \nu)^2 \right.\cr
\left. + {1\over2}\,\partial\nalpha\partial\beta \right] \, d V 
+ \int {1\over 2 r}\left( B - {A^2 \over B}\right) \cr
\times \left[ \partial_r \left( \nalpha - \beta/2 \right) 
+ {1\over r\,\tan\th} \partial_\th \left( \nalpha -
  \beta/2\right)\right] = 0\,.
\label{eq:GRV3}
\end{eqnarray}
Both of these virial theorems (\ref{eq:GRV2}) and (\ref{eq:GRV3}) can
be written as the sum of an integral over a ``material'' term $I_\mathrm{mat}$
(the first term in (\ref{eq:GRV2}) and (\ref{eq:GRV3}) respectively), 
and an integral over pure field-quantities $I_{\mathrm{fields}}$ (the
remaining terms). Therefore it will be convenient to consider the
following dimensionless quantity to numerically characterize the
respective virial violations:
\begin{equation}
  \label{eq:1}
  GRV \equiv {I_\mathrm{mat} + I_\mathrm{fields} \over I_\mathrm{mat}}\,.
\end{equation}

\subsection{The matter sources}

Let us write $\Gamma_\n$ and $\Gamma_\p$ for the two Lorentz factors
linking the Eulerian observers $\O_0$ to the co-moving fluid observers
$\O_\n$ (defined by $u_\n^\a$) and $\O_\p$ (defined by $u_\p^\a$), namely
\begin{equation}
\Gamma_\n \equiv - \eul_\a u_\n^\a  = N u_\n^t\,,\qaq
\Gamma_\p \equiv - \eul_\a u_\p^\a  = N u_\p^t\,.
\label{equGamma}
\end{equation}
The ``physical'' fluid velocities $U_\n$ and $U_\p$ of the two
fluids\footnote{In \citet{andersson01:_slowl_rotat_GR_superfl_NS}
  these were denoted $-\omega_\n$ and $-\omega_\p$ respectively.} 
in the $\varphi$ direction, as measured by $\O_0$, are given by 
\begin{equation}
U_\n = {1\over \Gamma_\n} \hat{\varphi}_\a u_\n^\a\,,\qaq
U_\p = {1\over \Gamma_\p} \hat{\varphi}_\a u_\p^\a\,,
\end{equation}
where $\hat{\varphi}^\a$ is the {\em spatial} unit vector in the
$\varphi$ direction, i.e. 
\begin{equation}
\hat{\varphi}^\a = {1\over\sqrt{g_{\varphi\varphi}}} \,\varphi^\a\,,
\quad\text{such\;\;that} \quad
h_{\a\b} \,\hat{\varphi}^\a \hat{\varphi}^\b = 1\,.
\end{equation}
Using (\ref{equ4Velocities}) and (\ref{equGamma}), we obtain
\begin{equation}
U_\n = {\sqrt{g_{\varphi\varphi}} \over N}\left( \Om_\n - N^\varphi \right),\;
U_\p = {\sqrt{g_{\varphi\varphi}} \over N}\left( \Om_\p - N^\varphi \right)\,,
\end{equation}
and the Lorentz factors can be expressed equivalently as
\begin{equation}
\Gamma_\n = \left(1 - U_\n^2 \right)^{-1/2}\,,\qaq
\Gamma_\p = \left(1 - U_\p^2 \right)^{-1/2}\,.
\end{equation}
The ``crossed'' scalar $x^2$, defined in (\ref{equScalars}), can be
expressed in terms of the respective scalar particle densities $n_\n$,
$n_\p$ and the 3--velocities $U_\n$ and $U_\p$, as
\begin{equation}
x^2 = n_\n n_\p {1 - U_\n U_\p \over \sqrt{(1-U_\n^2)(1- U_\p^2)}}\,,
\label{equx2}
\end{equation}
and using (\ref{eq:DefDelta}), we can write the relative velocity $\Delta$ as
\begin{equation}
\Delta^2 = { (U_\n - U_\p)^2 \over (1 - U_\n U_\p )^2 }\,.
\end{equation}
Using expressions (\ref{equFirstIntegral2a}),(\ref{equFirstIntegral2b})
and (\ref{equGamma}), the first integrals can be cast into the form 
\begin{eqnarray}
{N \over \Gamma_\n} \enth^\n = \text{const}^\n \,,\qaq
{N \over \Gamma_\p} \enth^\p = \text{const}^\p \,.
\label{equFirstIntegrals3}
\end{eqnarray}
In closer analogy with 
\cite{bonazzola93:_axisy,gourgoulhon99:_fast_rotation_strange_stars},
we can alternatively write these first integrals as
\begin{eqnarray}
H_\n + \nu - \ln \Gamma_\n &=& \C_\n\,,\\
H_\p + \nu - \ln \Gamma_\p &=& \C_\p\,,
\end{eqnarray}
where we introduced the abbreviations
\begin{equation}
H_\n \equiv \ln \left( \enth_\n \over m^\n \right)\,,\qaq 
H_\p \equiv \ln \left( \enth_\p \over m^\p \right)\,.
\end{equation}
The components of the $3+1$ decomposition (\ref{equ3+1}) of the
energy--momentum tensor (\ref{equTmunu}) are explicitly found as
\begin{eqnarray}
E &=&  - \Psi + (\Gamma_\n^2 \K^{\n\n} n_\n^2  + \Gamma_\p^2 \K^{\p\p} n_\p^2 \nonumber\\
&& \hspace{2cm} + 2 \Gamma_\n \Gamma_\p \K^{\n\p} n_\n n_\p ) \,,\label{eq:Matter1}\\
\sqrt{g_{\varphi\varphi}} \,J^\varphi &=&
\Gamma_\n^2 \K^{\n\n} n_\n^2 U_\n  + \Gamma_\p^2 \K^{\p\p} n_\p^2 U_\p \nonumber\\
&& \hspace{1cm} + \Gamma_\n \Gamma_\p \K^{\n\p} n_\n n_\p (U_\n + U_\p)\,,\\
S^r_r &=& S^\theta_\theta = \Psi\,,\\
S^\varphi_\varphi &=& \Psi + \left( \Gamma_\n^2 \K^{\n\n} n_\n^2 U_\n^2
  + \Gamma_\p^2 \K^{\p\p} n_\p^2 U_\p^2 \right. \nonumber\\
&& \hspace{1cm}\left. + 2\Gamma_\n \Gamma_\p \K^{\n\p} n_\n n_\p  U_\n U_\p \right)\,,
\label{eq:Matter4}
\end{eqnarray}
One can check the consistency of this result with the single
fluid case of \cite{bonazzola93:_axisy}, by considering the special
case of both fluids moving together.

\section{Numerical procedure}
\label{sec:numerical-procedure}

\subsection{Iteration scheme}
\label{sec:iteration-scheme}

The numerical solution of the stationary axisymmetric configurations
described in the previous sections proceeds in a very similar manner
to the single-fluid case, which is described in more detail in
\cite{bonazzola93:_axisy,gourgoulhon99:_fast_rotation_strange_stars}.
The central iteration scheme is nearly identical:\\

\begin{description}
\item[Initialization:] Start from a simple ``guess'' for a 
  spherically symmetric matter distribution $n_\n^\stat$ and
  $n_\p^\stat$ of the two fluids, and use a flat metric.  

\item[Step 1:] Calculate the matter source-terms $E$, $J^\varphi$ and $\S^i_j$ from
  (\ref{eq:Matter1})-- (\ref{eq:Matter4}).   

\item[Step 2:] Solve the equations (\ref{eq:Poisson3D})--(\ref{eq:Poisson2D}) 
  for the corresponding metric using the pseudo-spectral elliptic solver
  in LORENE, the numerical relativity package used here\citep{lorene}. 

\item[Step 3:] Use the first integrals (\ref{equFirstIntegrals3}) to
  obtain the chemical potentials $\enth^\n$ and $\enth^\p$. 

\item[Step 4:] Calculate the new density fields $n_\n$ and $n_\p$ by 
  inverting the relations (\ref{equEnthalpies}) for the given equation
  of state.

\item[Step 5:] Continue at Step~1  until the desired convergence is achieved.
\end{description}

In general we are using three spherical-type numerical domains to cover the
hypersurface $\Sigma_t$: the innermost domain covers the whole
star. An intermediate domain is used for the vicinity of the star to
about twice the stellar radius, and an outer domain covers the space
out to infinity, using a compactification of the type $u=1/r$ (see
\cite{bonazzola93:_axisy} for details). For the inner domain we have
the choice of either using a simple spherical grid containing the 
whole star, or we can use an adaptive-grid algorithm, in order to
adapt the domain-boundary to the stellar surface. Contrary to the 
single-fluid case (e.g. see \cite{gourgoulhon99:_fast_rotation_strange_stars}),
however, the adaptive-grid approach is much less effective in
increasing precision and convergence. The reason for this is simply 
that only one of the two fluid-surfaces can be matched up with a
domain boundary, and therefore the (weak) Gibbs phenomenon due to the
inner fluid surface (representing at least a discontinuity in the
derivative) is not completely avoidable.   

Another important difference in the case of two-fluids compared to
single-fluid stars is the way we determine the location of the fluid
surfaces. In a single fluid star, the surface can always be found
by the vanishing of the pressure, which usually translates into a simple
condition in terms of the vanishing of the chemical potential
$\mu$. In the two-fluid case, however, this is not generally possible
(especially for the ``inner'' fluid), due to the coupling of the
fluids. Therefore we need to define the fluid surfaces 
directly in terms of the vanishing of the respective density fields.
Contrary to the chemical potential, the density can have a vanishing or
diverging gradient at the surface, and a precise determination of the
surface can therefore be numerically difficult.\footnote{We note that
  \citet{yoshida04:_rapid_newton} chose to avoid this difficulty
  by \emph{defining} the ``fluid surfaces'' by the vanishing of the
  respective chemical potentials. These ``surfaces'', however, do
  generally \emph{not} coincide with the  surfaces of vanishing
  density (contrary to the single-fluid case), as can be seen from
  (\ref{equEnthalpies}).}  

A related numerical problem specific to two-fluid configurations
appears when the surfaces of the two fluids are very 
close to each other. In this case, the 1-fluid region in between the
two surfaces will be poorly resolved by the grid covering the star,
and therefore the determination of the outer surface will have a poor
numerical precision.
As will be seen later, this problem can be cured to some extent by
adding another domain, which covers just a thin shell below and up to
the outer fluid surface. In this case one observes a drastic
improvement in the precision of finding the outer surface, which can
be quantified by comparison with the analytic slow-rotation solution. 

We note that this numerical code can be used equally well for Newtonian
configurations, simply by replacing the
matter-sources by their Newtonian limits, and forcing the spatial
metric to be flat.
The central  iteration scheme remains unchanged, and we can relate the
lapse $N$ to the Newtonian gravitational potential, namely by the
relation $\nu = \ln N = \Phi/c^2$, where $\Phi$ is the Newtonian
gravitational potential. The Newtonian limit of the matter source-term
in Eq.~(\ref{eq:Poisson3D}) is 
\begin{equation}
{E + S_i^i \over c^2} = \rho + \O(c^{-2})\,,
\end{equation}
where $\rho$ is the total (rest-)mass-density, so that this component
of the Einstein equations reduces to the Newtonian Poisson equation,
while the remaining Einstein equations
(\ref{eq:PoissonVect})--(\ref{eq:Poisson2D}) become trivial in this
limit. In a similar manner, the first integrals are seen to reduce 
exactly to their Newtonian
counterparts~\cite{prix02:_slow_rot_ns_entrain}.

The parameters of the numerical scheme that will be used for the rest
of the paper are the following: the required convergence of the
iteration scheme is $10^{-10}$, and we use $17$ points in the $\theta$
direction, and $33$ grid-points in the radial direction in the
innermost domain (containing the star), $33$ radial points in the
intermediate domain and $17$ radial points in the compactified outer
domain. 

\subsection{The polytropic 2-fluid equation of state}

The numerical scheme described in the previous section can be used for
any invertible 2-fluid equation of state (EOS). The current
implementation of our code, however, only covers a ``polytropic'' 
subclass of 2-fluid EOS, which generalizes the types of EOS used in
previous studies, e.g.
\cite{prix02:_adiab,prix02:_slow_rot_ns_entrain,andersson01:_slowl_rotat_GR_superfl_NS,yoshida04:_rapid_newton},
and which has the general form
\begin{eqnarray}
\E &=& \rho \,c^2 + 
{1\over2} \kapn \n^{\gamma_{1}} + 
{1\over2} \kapp n_\p^{\gamma_{2}} 
+ \kapnp n_\n^{\gamma_{3}} n_\p^{\gamma_{4}} \nonumber\\
&& + \kapD  n_\n^{\gamma_{5}} n_\p^{\gamma_{6}} \Delta^2\,,
\label{eq:EOS}
\end{eqnarray}
where $\rho \equiv m_\n n_\n + m_\p n_\p$.  
As discussed in the introduction, we expect this polytropic EOS-class 
to be quite general, and to allow one to study
the qualitative features of a broad range of different superfluid 
neutron star models.  For example, general features of the Kepler limit 
(cf.~Fig.~\ref{fig:Kepler}) are seen to be in qualitative agreement with 
the mean field results of \citet{comer04:_slowl}. 

The two fluids in~(\ref{eq:EOS}) are coupled via a ``symmetry energy''-type
term proportional to $\kappa_{\n\p}$ and an entrainment term
proportional to $\kapD \Delta^2$.
The resulting expressions for the chemical potentials and the
entrainment $\entr$ are directly obtainable from (\ref{equFirstLawII}). 

In general this class of 2-fluid EOS requires a numerical inversion 
in the iteration scheme described in section~\ref{sec:iteration-scheme}, 
in order to obtain the densities $n_\n, n_\p$ from the chemical
potentials $\enth^\n, \enth^\p$ at a given relative speed $\Delta$.
For testing purposes and for comparison to the Newtonian and
relativistic slow-rotation results, we will in the following be
mostly interested in a further subclass of the above EOS, namely
the special 2-fluid polytropes described by 
\begin{equation}
\E =  \rho\,c^2 + {1\over 2} \kapn n_\n^2 + {1\over 2} \kapp n_\p^2 +
\kapnp n_\n n_\p + \kapD n_\n n_\p \Delta^2\,,
\label{eq:anEOS}
\end{equation}
which are a 2-fluid generalization of the 1-fluid polytrope $P\propto n^2$.  
This special EOS class still exhibits all the coupling-types
(entrainment + symmetry energy) of the general EOS, but allows an
analytic inversion, namely
\begin{eqnarray}
\enth^\n - m_\n c^2 &=& \kapn n_\n + (\kapnp + \kapD \Delta^2) \, n_\p \,,\\
\enth^\p - m_\p c^2 &=& \kapp n_\p + (\kapnp + \kapD \Delta^2) \, n_\n \,,
\end{eqnarray}
and the entrainment is found as
\begin{equation}
\entr = \kapD n_\n n_\p \,.
\end{equation}
The generalized pressure $\Psi$ in (\ref{equPressure2}) is expressible as 
\begin{equation}
\Psi = {1\over2} \kapn \n^2 + {1\over2} \kappa_\p n_\p^2 
+ \kappa_{\n\p} n_\n n_\p + \kapD n_\n n_\p \Delta^2\,.
\end{equation}
Contrary to the two-fluid EOS used in the Newtonian slow-rotation
study \cite{prix02:_slow_rot_ns_entrain}, which exhibits the somewhat
unphysical feature of constant entrainment numbers, as discussed in
appendix~\ref{sec:newt-analyt-slow}, this EOS results in a much more 
physical behavior of the entrainment. Namely, using~(\ref{eq:10}), we find
\begin{equation}
\eps_\n = {2 \kapD \over m^\n} \,n_\p\,, \qaq \eps_\p = {2 \kapD \over m^\p}\, n_\n\,,
\end{equation}
which ensures that the entrainment effect automatically vanishes when
one of the two fluid-densities goes to zero.
Such a linear behavior of entrainment also happens to be in quite good
qualitative agreement with the theoretical predictions of nuclear 
physics, e.g. see\footnote{These references give the neutron and proton
  effective masses $m^{\X*}$, which are related to
  the entrainment via $\eps_\X = (m^\X-m^{\X*})/m^\X$, see
  \cite{prix02:_slow_rot_ns_entrain} for details.} 
\cite{chao72:_proton_superfl,sjoberg76:_effect_mass,baldo92:_superfl_neutr_star_matter}

Using the method developed in~\cite{prix02:_slow_rot_ns_entrain}
for the EOS (\ref{eq:anEOS}), we can find an analytic solution in the
Newtonian slow-rotation approach, which is presented in
appendix~\ref{sec:newt-analyt-slow}.   
This allows us to run extensive tests by comparing our numerical
code to the analytic solution in the Newtonian case. The results of
this comparison are presented in section~\ref{sec:comp-newt-slow}.

\section{Tests of the numerical code}
\label{sec:tests-numerical-code}

\subsection{Comparison to 1-fluid results}

\begin{figure*}[htbp]
  \psfrag{dM}{$\Delta M$}
  \psfrag{dRpole}{$\Delta R_\pol$}
  \psfrag{dReq}{$\Delta R_\equ$}
  \psfrag{Omega}{$\Om / \Om_0$}
  \psfrag{fixed grid}{fixed grid}
  \psfrag{adaptive grid}{adaptive grid}
  \psfrag{(a)}{(a)}
  \psfrag{(b)}{(b)}
  \psfrag{(c)}{(c)}
  \psfrag{(d)}{(d)}
  \psfrag{Newtonian}{Newtonian}
  \psfrag{relativistic}{relativistic}
  \centering
  \includegraphics[width=0.8\textwidth,clip]{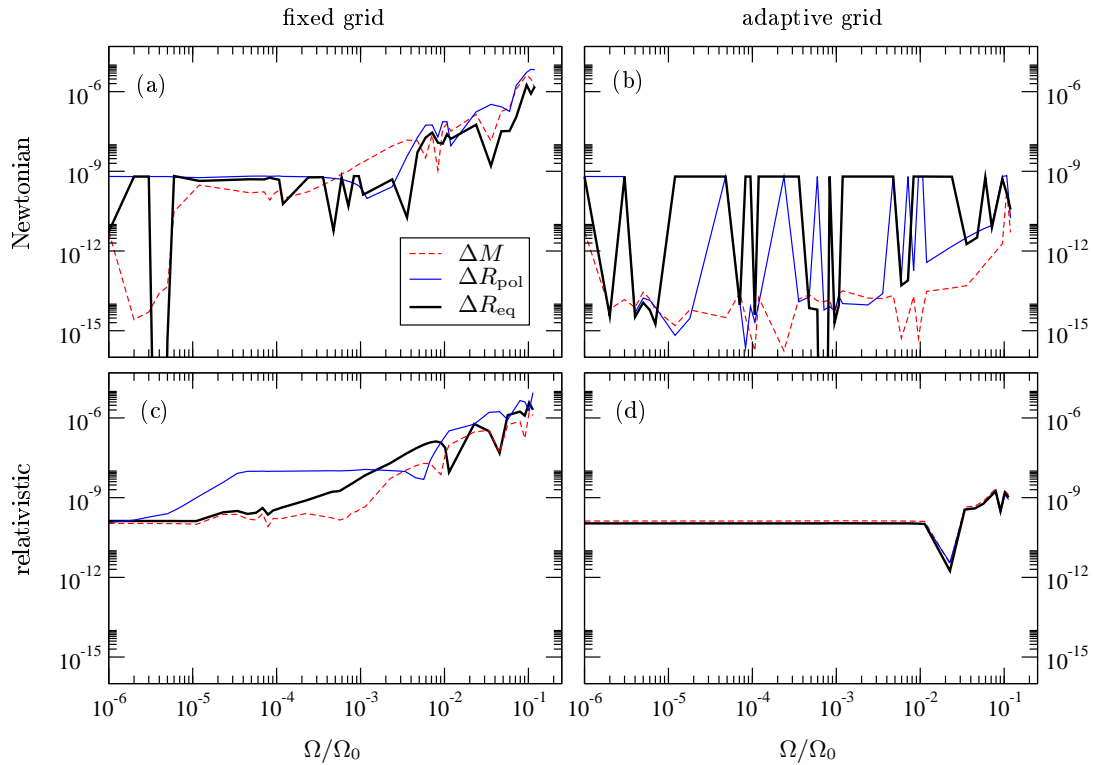}
  \caption{Relative differences $\Delta Q$ in the total baryon mass
    $M$, and the equatorial and polar radii 
    $R_\equ$ and $R_\pol$. The first row is the Newtonian case,
    while the second row shows the relativistic results. 
    In the first column, we used a fixed spherical
    inner domain, while in the second column the inner domain-grid is
    adapted to the stellar surface (which is the default in the 
    single-fluid case).}
\label{fig:Comp1f}
\end{figure*}

As a first consistency check we use the two-fluid code for  
strictly co-rotating configurations with a common outer surface, and
compare the results to those of the well-tested single-fluid code 
\cite{bonazzola93:_axisy,gourgoulhon99:_fast_rotation_strange_stars}.
For this purpose we study a stellar sequence of fixed central density
and vary the rotation rate. We define the ``natural scale'' of the
rotation-rate as 
\begin{equation}
  \label{eq:2}
  \Om_0 \equiv \sqrt{ 4\pi G \, \rho(0)}\,,
\end{equation}
where $\rho(0)$ is the central rest-mass density, i.e. 
$\rho(0) = m^\n n_\n(0) + m^\p n_\p(0)$. The Kepler rotation rate 
$\Om_\Kepler$ is typically found at about $\Om_\Kepler \sim 0.1\,\Om_0$ 
for the configurations considered here.
The results of the comparison with the single-fluid case are shown in
Fig.~\ref{fig:Comp1f}. Here we plot the relative differences, defined as
\begin{equation}
  \label{eq:3}
  \Delta Q \equiv { |Q^\tf - Q^\of | \over Q^\of}\,,
\end{equation}
of a global quantity $Q$ in the two-fluid case ($Q^\tf$) and in the
single-fluid case ($Q^\of$).
The first column, figures~\ref{fig:Comp1f}~(a) and (c), shows the
comparison of 1-fluid and 2-fluid results using a \emph{fixed}
spherical grid for the inner domain in the two-fluid case.
The single-fluid code on the other hand always uses an adaptive
grid for the stellar surface.
We notice that towards  higher rotation rates the relative errors
increase. These errors can be entirely ascribed to the lack
of grid adaption in the two-fluid case: by using an adaptive grid for
the stellar domain also in the two-fluid case, we find a consistent
agreement of better than $10^{-9}$, as can be seen in 
the second column in figure~\ref{fig:Comp1f}~(b) and (d). 
We note, however, that this improvement of using an adaptive grid is
restricted to cases where the two-fluids share a common outer surface,
while it is of much less use in the general two-fluid case as
mentioned earlier.
We can therefore conclude that the two-fluid code reproduces results
consistent with the single-fluid code in cases where the two fluids
co-rotate. 

\subsection{Virial theorem violation}

In the next step we consider the more general case where the two
fluids have different rotation rates. We fix the relative rotation
rate, defined as  
\begin{equation}
\R \equiv {\Om_\n - \Om_\p \over \Om_\p}\,,
\end{equation} 
to be $\R=1.51$ and vary $\Om_\n$. As mentioned before, in these
general situations an adaptive grid does not substantially
improve the precision and has therefore not been used.
\begin{figure}[htbp]
  \psfrag{Omega_n}{$\Om_\n / \Om_0$}
  \psfrag{OmK}{$\Om_\Kepler$}
  \psfrag{GRV}{GRV}
  \centering
  \includegraphics[width=0.5\textwidth,clip]{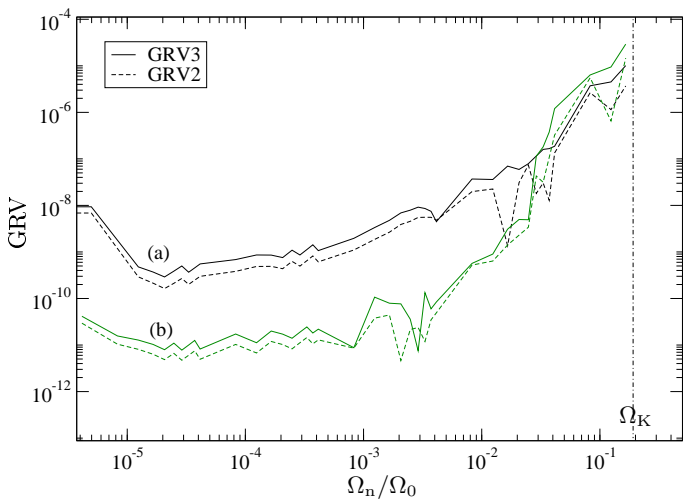}
  \caption{Virial violations GRV2 and GRV3 for (a) 1 domain and (b) 2
    domains covering the star. Only the Newtonian case is shown, as
    the relativistic results are very similar. $\Om_\Kepler$ denotes
    the Kepler rotation rate.}
  \label{fig:GRVs}
\end{figure}
We consider the internal consistency check provided by the virial
identities GRV2 and GRV3 defined in~(\ref{eq:1}), for which the result 
is shown in Fig.~\ref{fig:GRVs}.  We note that in the case (a), where
one inner domain is used to cover the star, even at low rotation rates
the result falls somewhat short of the convergence-goal of $10^{-10}$
in the iteration scheme. This lack of precision at
small rotation rates can be understood as follows: due to the
difference in rotation rates, the two fluids do not share a common
outer surface, and there will necessarily be a 1-fluid region close to
the outer surface. However, this 1-fluid region will be very thin
compared to the dimensions of the star, and will therefore be poorly
resolved in terms of the numerical grid. We can improve this by
choosing a second domain to cover just a thin layer (of about $1\%$ of
the radius) below the outer surface, resolved by another $33$ radial
grid-points. The effect of this ``trick'' is rather impressive and can
be seen in Fig.~\ref{fig:GRVs}, for the case (b).  While this gain in
precision is not very important by itself, it underlines the
consistency of the results and shows that the source of these errors
is understood.  The decrease in precision when approaching the Kepler
rotation can be ascribed to the appearance of cusps at the
equator (see Fig.~\ref{fig:KeplerStars}) and therefore the presence of
the Gibbs phenomenon. Nevertheless, one should note that this
phenomenon happens also in the one-fluid case, the precision of the
code at the Kepler limit being of the same order as here
\cite{gourgoulhon99:_fast_rotation_strange_stars}.

\subsection{Comparison to Newtonian slow-rotation results}
\label{sec:comp-newt-slow}

We can use the analytic Newtonian solution in the slow-rotation
approximation (derived in appendix~\ref{sec:newt-analyt-slow})
for a systematic comparison with the numerical code run in ``Newtonian
mode'' as described in section~\ref{sec:iteration-scheme}.

We denote the numerical solution of a quantity as $Q_\L(\Om_\X)$,
and the analytic slow-rotation solution as $Q_\sr(\Om_X)$, and we
define the relative difference as
\begin{equation}
\rel Q \equiv {Q_\L - Q_\sr \over Q^\stat }\,,
\label{eq:DefRelQ}
\end{equation}
where $Q^\stat$ corresponds to the static solution.
At fixed relative rotation rate $\R$, the slow-rotation solution can
be written as  
\begin{equation}
Q_\sr(\Om_\n) = Q^\stat + Q^\snd\, \Om_\n^2\,.
\label{eq:SolSlowRot}
\end{equation}
Ideally we would like to compare only up to the $\Om^2$ component of the
numerical solution, but obviously we do not know its Taylor-expansion
in orders of $\Om$. Nevertheless the numerical solution can formally
be written as
\begin{equation}
Q_\L(\Om_\n) = Q_\L^\stat + Q_\L^\snd\, \Om_\n^2 + Q_\L^\fth\, \Om_\n^4 + ...\,,
\label{eq:SolLorene}
\end{equation}
so that the relative difference~(\ref{eq:DefRelQ}) can be expanded as
\begin{equation}
\rel Q = {Q_\L^\stat - Q^\stat \over Q^\stat} 
+ {Q_\L^\snd - Q^\snd \over Q^\stat} \, \Om_\n^2
+ {Q_\L^\fth \over Q^\stat}\, \Om_\n^4 + ...\,.
\end{equation}
If the numerical solution agreed perfectly with the analytic solution
(up to order $\Om^2$), the first two terms would be zero, and the
leading order of the difference would be $\Om_\n^4$. In practice,
however, there will be contributions on all orders, and we will try to
quantify these respective errors. If in some interval of $\Om_\n$ one
of these terms dominates in the series, then a log-log plot of $\rel
Q(\Om_\n)$ in this region would look like 
\begin{equation}
y = \log(a\, \Om_\n^m) = \log a + m \, \log\Om_\n\,,
\end{equation}
i.e. a straight line with steepness $m$ and an offset $\log a$.
Conversely, if the log-log plot of $\rel Q$ contains sections of
straight lines, we can infer the leading power in $\Om_\n$ 
and its coefficient.

The neutron-star model used here is characterized by the following
choice of EOS-parameters:  
\begin{equation}
\kapn = 0.02\,,\;
\kapp = 0.12\,,\;
\kapnp = 0.01\,,\;
\kapD = 0.02\,,
\end{equation}
and the (fixed) central chemical potentials
\begin{equation}
\enth^\n(0) = \enth^\p(0) = 0.2 \, m_b\, c^2\,,
\end{equation}
where $m_b = 1.66\times10^{-27}$~kg is the baryon mass and $c$ is the
speed of light.
The resulting neutron-star model in the static case has a total 
mass of $M = 1.50\,M_\odot$ (where $M_\odot$ is the solar mass), a
radius of $R = 11.1$~km, a central baryon number density $\n(0) =
1.04\;\fmq$, proton fraction $\xp \equiv n_\p/n = 0.083$ and an
entrainment value of $\eps_\p(0)= 0.38$.
We note that in the Newtonian case there is no distinction between the 
gravitational mass $\Mg$ and the baryon mass $M$.

\begin{figure*}[htbp]
  \psfrag{Rn_equ}{$R_\n(\equ)$}
  \psfrag{Rp_equ}{$R_\p(\equ)$}
  \psfrag{Om2}{$\Om^2$}
  \psfrag{Om4}{$\Om^4$}
  \psfrag{log10dQ}{$\log_{10}(\rel R)$}
  \psfrag{log10Om_n}{$\log_{10}(\Om_\n / \Om_0)$}
  \psfrag{(a)}{(a)}
  \psfrag{(b)}{(b)}
  \centering
  \includegraphics[width=0.8\textwidth,clip]{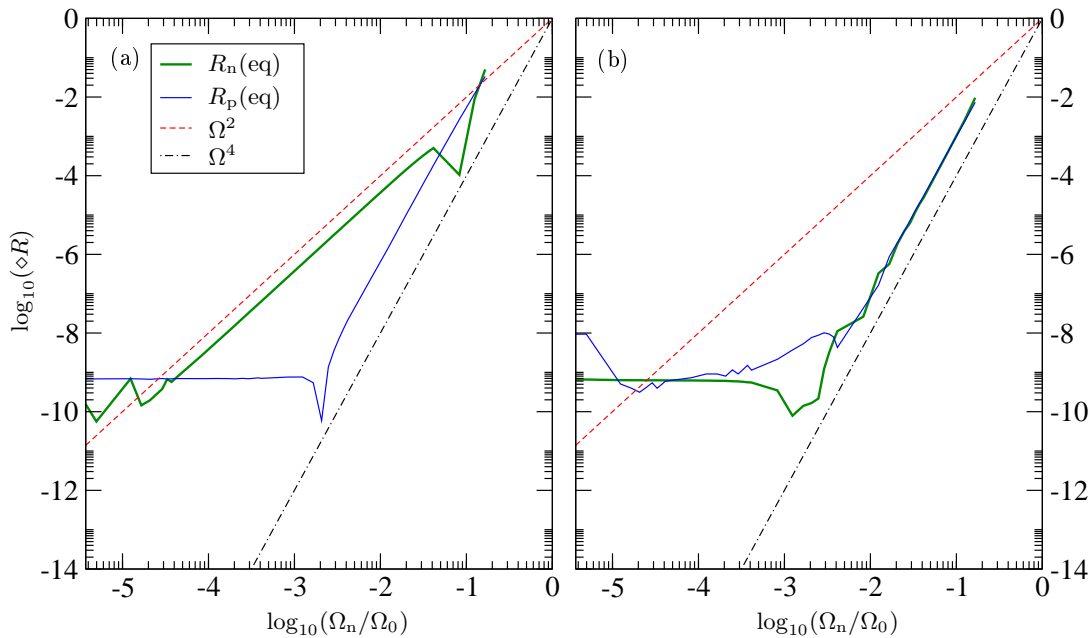}
  \caption{Relative difference $\rel Q (\Om_\n)$ between the numerical
  code in  ``Newtonian mode'' and the  slow-rotation analytic solution of
  appendix~\ref{sec:newt-analyt-slow}, for the equatorial radii $R_\n$
  and $R_\p$. In (a) we used the normal ``physical'' EOS-inversion,
  while (b) shows the results when using a ``slow-rotation style'' EOS inversion.}
  \label{fig:CompConv}
\end{figure*}

In Fig.~\ref{fig:CompConv} we show the relative differences $\rel Q$
for the radii $R_\n$ and $R_\p$ at the equator.
We also plotted the straight lines corresponding to a pure $\Om^4$
and $\Om^2$ behavior, in order to facilitate the interpretation of
these results.  

In Fig.~\ref{fig:CompConv}~(a) we see that for small
rotation rates ($\Om_\n / \Om_0< 10^{-3}$) the error 
in the equatorial proton radius $R_\p(\equ)$ reaches a ``plateau'' at
about $\sim 10^{-9}$, which corresponds to numerical errors and the
finite convergence-condition of the iteration scheme, while  
for higher rotation rates, the quartic error starts to dominate.
The same behavior is observed for other global quantities,
e.g. $M_\n$ and $M_\p$ and $R(\pol)$), which are not included 
in this plot. However, the neutron equatorial radius
$R_\n(\equ)$ (which is the outer radius) displays a consistent
\emph{quadratic} error of order unity!    
The reason for this apparent discrepancy is rather subtle, and stems
from the somewhat different nature of the slow-rotation approach and
the fully numerical solution. In the numerical code, when one of the
two fluid-densities vanishes, we switch from the 2-fluid EOS
(\ref{eq:EOS}) to the corresponding 1-fluid EOS \emph{before} we do
the inversion $\enth^\X \rightarrow n_\Y$ in the numerical procedure
(cf. section~\ref{sec:iteration-scheme}), which is the correct
physical way to do this.
In the slow-rotation approach, however, the rotation rates
are treated as infinitesimal, and there is actually no finite 1-fluid
region. Therefore the EOS is always used in the form (\ref{eq:EOS}),
which will be seen in the following to account for the difference in
$R_\n(\equ)$. In order to test this explanation, we have also
implemented a ``slow-rotation style'' EOS-inversion in the code, in
which we \emph{do not} switch to a 1-fluid EOS when one of the two
fluids vanishes. The result of this is shown in
Fig.~\ref{fig:CompConv}~(b). We see that the discrepancy of the
outer radius has completely disappeared. 
While this serves as an interesting test of consistency, this rather
unphysical EOS-inversion will obviously not be used in the following. 

\subsection{Comparison to relativistic slow-rotation results}

Finally, we compare our results in the fully relativistic case to the 
results obtained by using a code developed by
\citet{andersson01:_slowl_rotat_GR_superfl_NS}, which is based on the 
relativistic slow-rotation approximation. 

In the relativistic case the physical ``radius'' will generally be
different from the coordinate-radius, and can be defined in
various non-equivalent ways (e.g. circumferential radius, proper radius). 
For an unambiguous comparison we define the ``radius'' $R$ as the
proper distance of the surface from the center of the star, along a
line of constant $\theta$ and $\varphi$ (the definition of which is
consistent with \cite{andersson01:_slowl_rotat_GR_superfl_NS}), i.e.
\begin{equation}
  \label{eq:5}
  R \equiv \int_0^{R_0} d\,l = \int_0^{R_0} A(r)\, d\,r\,,
\end{equation}
where $R_0$ is the coordinate-radius of the surface.
Another quantity specific to the relativistic case is the shift-vector
$N^i = (0, 0, N^\ph)$, and we will consider its 3-norm, i.e. 
\begin{equation}
  \label{eq:4}
  ||N^i||\equiv \sqrt{g_{i j}N^i N^j} =  N^\ph\, \sqrt{g_{\ph \ph}} \,,
\end{equation}
which is independent of the coordinate-system chosen on the spacelike
hypersurface.

The stellar model used in this comparison is defined by  the EOS
parameters  
\begin{equation}
\kapn = 0.04\,,\;
\kapp = 0.24\,,\;
\kapnp = 0.02\,,\;
\kapD   = 0.02\,,
\label{eq:EOS2}
\end{equation}
and the central chemical potentials are $\mu^\n(0) = \mu^\p(0) =
0.2\,m_b\,c^2$. The configurations obtained have the following (fixed)
central values: the central baryon density is $\n(0) =
0.5776\;\fmq$, which corresponds to $3.61$ times nuclear density
($n_\nuc=0.16\;\fmq$). The central proton entrainment is $\eps_\p(0) =
0.212$, and the proton fraction is found as $\xp(0) = 0.083$.  
We fix the relative rotation to $\R=0.5$, i.e. the neutron superfluid
is rotating $50\%$ faster than the proton-electron fluid.
\begin{table}[htb]
  \begin{tabular}{ | c || c | c | c | }\hline
$\Om_\n / 2\pi$         &  0 Hz    &  100 Hz         &  500 Hz         \\ \hline\hline
$M_\n\,[\Msol]$         &  1.0978 (-0.02\%) &  1.0998 (-0.1\%) &  1.1509 (-2\%)  \\
$M_\p\,[\Msol]$         &  0.0998 (-0.02\%) &  0.0997 (-0.1\%) &  0.0959 (-2\%)  \\
$\Mg \,[\Msol]$         &  1.1194 (-0.02\%) &  1.1210 (-0.04\%) & 1.1644 (0.04\%) \\
$R_\n^\equ\,[\textrm{km}]$ & 13.545 (-0.01\%) &  13.570 (0.2\%) &  14.260 (5\%) \\
$R_\n^\pol\,[\textrm{km}]$ &  13.545 (-0.01\%) &  13.527 (-0.1\%) &  13.103 (-3\%) \\
$R_\p^\equ\,[\textrm{km}]$ &  13.545 (-0.01\%) &  13.534 (-0.1\%) &  13.302 (-2\%) \\
$R_\p^\pol\,[\textrm{km}]$ &  13.545 (-0.01\%) &  13.527 (-0.1\%) &  13.103 (-3\%) \\
$N(0)$                  &  0.700102 (1e-4\%) &  0.69983 (2e-3\%) &  0.69267(-0.04\%) \\
$||N^i||(\equ)$         &  0            &  0.00206 (0.1\%) & 0.01072 (3\%) \\\hline
\end{tabular}\\[0.2cm]
\caption{Numerical results $Q_\L$ and (in parentheses) relative
  differences  $(Q_\L - Q_\sr)/Q_\L\times 100\%$ to the  
  relativistic slow-rotation results $Q_\sr$. 
} 
\label{tab:CompareGreg}
\end{table}
In Table~\ref{tab:CompareGreg} we show the results of the comparison to
the relativistic slow-rotation code. We observe that generally the
agreement is quite good, and (as expected) gets worse with higher
rotation rates. However, we note that this slow-rotation code imposes
an additional constraint on the radii, namely the two fluids are
forced to share a common outer surface. Therefore part of the
disagreement observed here does not actually stem from the slow-rotation
approximation or numerical differences, but from the somewhat different
assumptions in the model. Given these differences, the agreement seems
very good.  

\section{Numerical Results}
\label{sec:numerical-results}

The existence of configurations with one fluid-surface having a
prolate shape was initially found using the Newtonian analytic
solution\cite{prix02:_slow_rot_ns_entrain} in the slow-rotation
approximation. While this might not be very realistic astrophysically, 
it is still interesting to study this particularity of an interacting
two-fluid system. We confirm the existence of such configurations
in the fully relativistic  treatment, as reported earlier by
us\cite{prix02:_relativistic_sf_ns}.  
In order to show this, we choose the polytropic EOS parameters $\kapn
= 0.016$, $\kapp = 0.16$, $\kapnp = 0.008$, and $\kapD = 0.03$, with
the central chemical potentials $\mu^\n(0) = 0.2 m_b c^2$ and
$\mu_\p(0) = 0.198 m_b c^2$. This corresponds to a central proton
fraction of $\xp(0) = 0.05$ and a central proton-entrainment number of
$\eps_\p(0) = 0.80$. 
\begin{figure}[htbp]
  \psfrag{Om}{$\vec{\Omega}$}
  \psfrag{x [km]}{$x$ [km]}
  \psfrag{z [km]}{$z$ [km]}
  \centering
  \includegraphics[width=0.4\textwidth,clip]{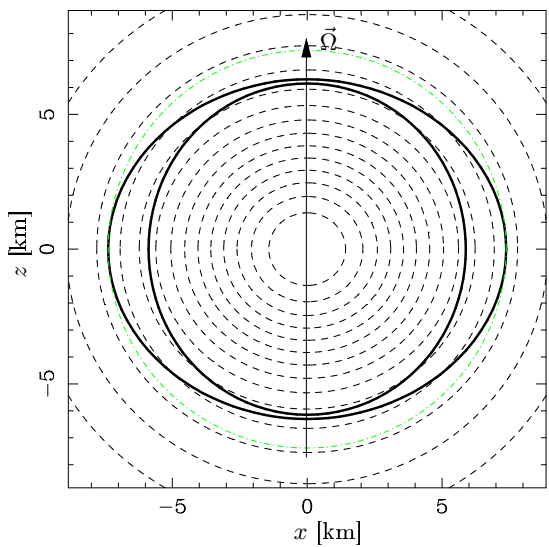}
  \caption{Meridional cross-section of an  oblate-prolate
    two-fluid configuration. The dotted lines represent lines of
  constant ``gravitational potential'' $N$, while the thick lines are
  the respective surfaces of the neutron-  and proton fluids.}
\label{fig:prolate}
\end{figure}
\begin{figure*}[htbp]
  \psfrag{OmK}{$\Om_\Kepler [s^{-1}]$}
  \psfrag{Rel}{$\R$}
  \psfrag{Om_n}{$\Om_\n$}
  \psfrag{Om_p}{$\Om_\p$}
  \psfrag{analytic}{NSR}
  \psfrag{model_I}{EOS~I}
  \psfrag{model_II}{EOS~II}
  \psfrag{model_III}{EOS~III}
  \centering
  \includegraphics[width=0.9\textwidth,clip]{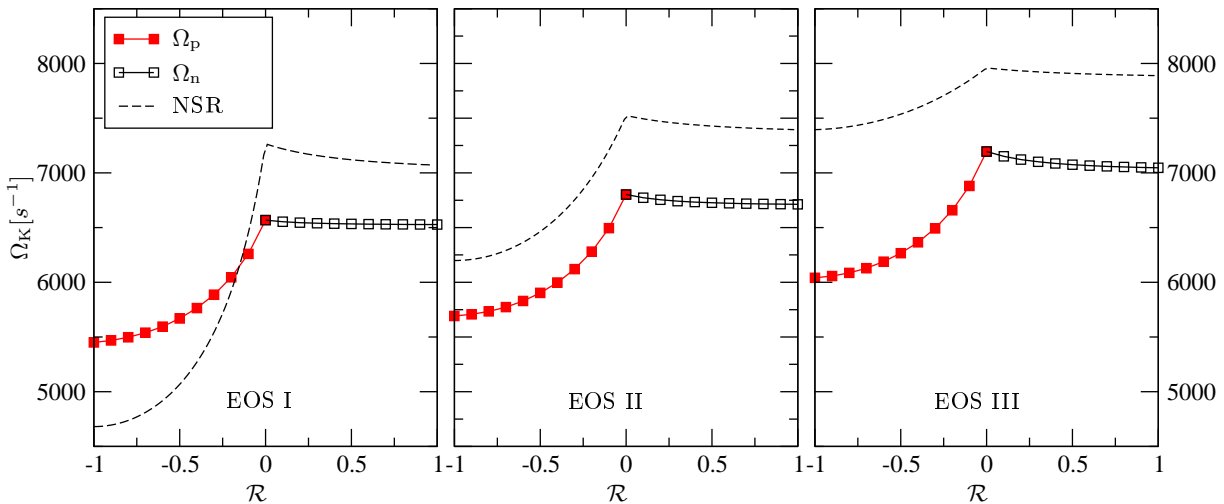}
  \caption{Kepler limit $\Om_\Kepler$ as a   function of relative
    rotation rate $\R$ for EOS-models~I, II and III. The dashed line
    (NSR) represents the result from the analytic Newtonian
    slow-rotation solution (cf.~Appendix~\ref{sec:newt-analyt-slow})}   
  \label{fig:Kepler}
\end{figure*}
In Fig.~\ref{fig:prolate} we show the resulting
configuration with the two fluids counter-rotating at 
$\Om_\n/2\pi = 1000$~Hz and $\Om_\p/2\pi = -100$~Hz. 
We define the ellipticity of fluid $\X$ as 
\begin{equation}
\epsilon_\X \equiv {R_\X(\equ) - R_\X(\pol) \over  R_\X(\equ)} \,,
\label{eq:8}
\end{equation}
in terms of the proper radii $R$ of~(\ref{eq:5}). 
Using this definition, this configuration is found to have
$\epsilon_\n = 0.137$, and $\epsilon_\p = -0.037$, so the proton fluid
has a prolate shape despite the fact that it is rotating around the z-axis. 
This is made possible by the effective interaction potential created
by the neutron-fluid, which ``squeezes'' the proton-fluid, in this
case to the point of even overcoming the centrifugal potential.
\begin{table}[htb]
  \centering
  \begin{tabular}{| c || c | c | c | c |}\hline
    EOS &  $\kapn$  & $\kapp$  & $\kapnp$  & $\kapD$ \\\hline
     I  &  $0.05$   &  $0.5$  &  $0.025$   & $0.02$  \\
     II&   $0.05$   &  $0.5$  &  $0.0$     & $0.0$   \\
     III&  $0.05$   &  $0.5$  &  $-0.025$  & $0.02$ \\\hline
  \end{tabular}\\[0.2cm]
  \caption{Polytropic parameters defining EOS-models~I, II, and III}
  \label{tab:DefineModels}
\end{table}

\begin{figure}[htbp]
  \psfrag{MassG}{$\Mg$}
  \psfrag{nbar_fm-3}{$n_c$~[$\fmq$]}
  \psfrag{Kepler_R0}{$\R=0$}
  \psfrag{Kepler_R0.5}{$\R=0.5$}
  \psfrag{Kepler_R-0.5__}{$\R=-0.5$}
  \centering
  \includegraphics[width=0.5\textwidth,clip]{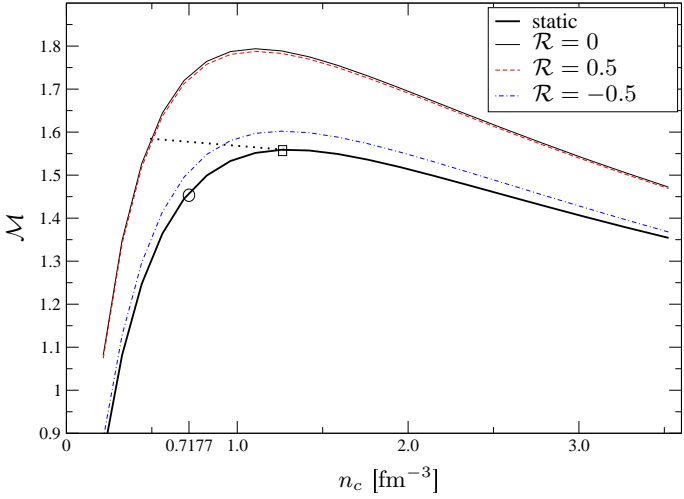}
  \caption{Gravitational mass $\Mg$ as a function of central baryon
    number density $n_c$ for EOS~I with fixed central proton
    fraction of $\xp=0.05$. The four curves correspond to the
    non-rotating case, the co-rotating Kepler-configuration
    ($\R=0$), and two Kepler-configurations with relative rotation rates
    of $\R=0.5$ and $\R=-0.5$ respectively.
    The circle indicates a static configuration with central
    chemical potentials of $\mu_\n=\mu_\p =0.3\;m_b c^2$. The box
    indicates the maximum-mass configuration in the static case. The
    dotted line represents the sequence of constant baryon mass 
    connecting to the static maximum-mass configuration. 
    Configurations above this line have no stable non-rotating
    counterpart and are called ``supramassive'' stars.}  
  \label{fig:MassDens}
\end{figure}

To simplify the presentation of results, we focus in the following on
three EOS-models, defined in table~\ref{tab:DefineModels}, which
differ only by their interaction-terms.   
The EOS-models~I and III differ by the sign of the
``symmetry-interaction'' term $\kapnp$, which corresponds to a value
of the canonical ``symmetry-energy term'' (introduced in
\citet{prix02:_slow_rot_ns_entrain}) of $\sigma=-0.5$ for EOS~I and
$\sigma=0.5$ for EOS~III. EOS~II represents two fluids without
EOS-interactions.
If not otherwise stated, we choose the central chemical potentials as
$\mu_\n = \mu_\p = 0.3\,m_b\,c^2$. In the static case we obtain the
results shown in Table~\ref{tab:CompareModels} for these three
EOS-models. We note that all three static configurations are on the
stable branch of the mass-density relation, which can be seen
in Fig.~\ref{fig:MassDens} for EOS-model~I.
\begin{table}[htb]
  \centering
  \begin{tabular}{| c | c | c | c |}\hline
                        &    EOS I    &  EOS II      &   EOS III\\\hline
   $n_c\;[\fmq]$  &    0.7177      &  0.7697      & 0.8612      \\
   $\eps_\p(0)$          &    0.273      &  0.0        & 0.301      \\
   $\xp(0)$              &    0.05       &  0.09       & 0.125      \\
   $M\;[ \Msol]$         &    1.586      &  1.532      & 1.448      \\
   $\Mg\; [\Msol]$       &    1.460      &  1.409      & 1.332      \\
   $R\; [\mathrm{km}]$   &    14.37      &  13.88      & 13.12      \\\hline
   \end{tabular}\\[0.2cm]
   \caption{Results for the central baryon number density $n_c$, entrainment
   $\eps_\p$, proton fraction $\xp$, total baryon mass $M$,
   gravitational mass $\Mg$ and the proper radius $R$ for
   EOS-models~I, II and III in the static case.}  
   \label{tab:CompareModels}
\end{table}
We note that when considering rotation, the individual fluid radii
will obviously change, but also the masses, because we only consider
stellar sequences of fixed central density.   

Next we consider these stellar models rotating at their maximum rotation
rate $\Om_\Kepler$ (called Kepler limit) for different relative
rotation rates $\R$, as shown in Fig.~\ref{fig:Kepler}. We define
the Kepler-rate $\Om_\Kepler$ as the rotation rate of the outer
fluid (which in this case also happens to be the faster rotating one),
i.e. the protons for $\R<0$ and the neutrons for $\R>0$. The rotation
rate of the (slower) inner fluid is trivially determined by $\Om_\Kepler$ and
$\R$. The dashed line shows the result from the Newtonian
slow-rotation solution (\ref{eq:15}). This is seen to overestimates
the Kepler rate typically by about $15\,\%$, except for 
the case of EOS~I, where it can even underestimate the Kepler limit
for $\R<0$. We see that in the fixed
central-density sequences considered here, the local maximum of the
Kepler rate is always attained  for the co-rotating configuration
(i.e.~$\R=0$), which contrasts with the case of fixed-mass sequences
as considered in the Newtonian study \cite{prix02:_slow_rot_ns_entrain}.  
A similar feature of the Kepler-rate decreasing as $\R$ decreases through zero 
can be seen in the mean-field results of \citet{comer04:_slowl}.

\begin{figure*}[htbp]
  \psfrag{(a)}{(a)}
  \psfrag{(b)}{(b)}
  \psfrag{Om}{$\vec{\Omega}$}
  \centering
  \includegraphics[width=0.8\textwidth,clip]{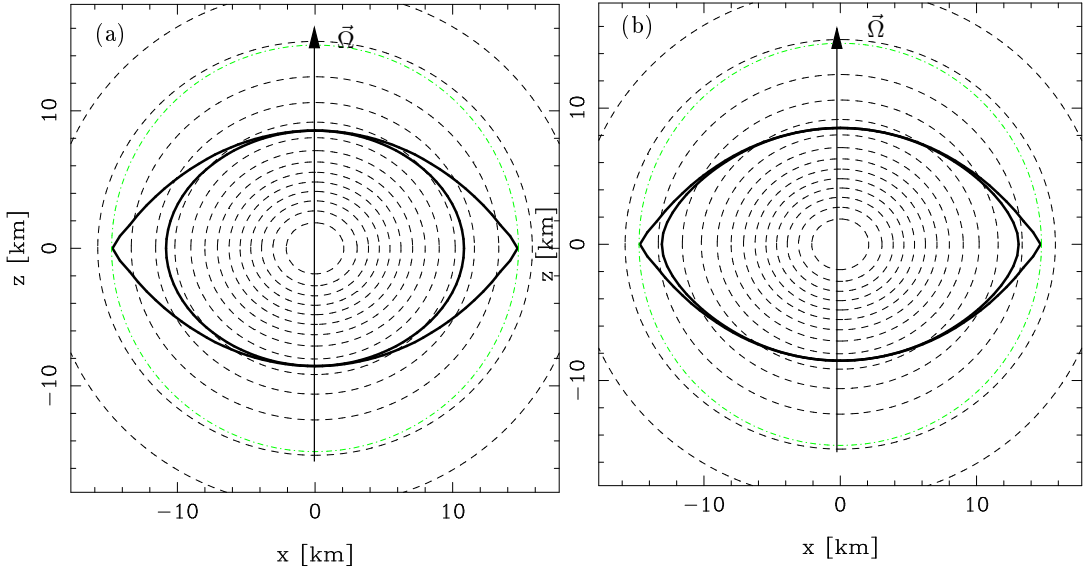}
  \caption{Kepler configurations for EOS~I. In figure (a) the relative
    rotation rate is $\R=0.1$, while in (b) it is $\R=0.01$.}
  \label{fig:KeplerStars}
\end{figure*}

In the Fig.~\ref{fig:KeplerStars}, we show the fluid
surfaces of the two fluids rotating at the Kepler-rate for two
different relative rotation rates, $\R=0.1$ and $\R=0.01$
respectively. We see the characteristic ``cusp'' appearing at the
equator of the outer fluid, which indicates the onset of
mass-shedding if the rotation rate were to be increased any further. Because
we fixed the central densities of these configurations to those of
the static case, it can be seen from Fig.~\ref{fig:MassDens} that both
of these configurations belong to the so-called ``supramassive''
class, i.e. they do not have a corresponding stable non-rotating
configuration of equal baryon-mass. 

Fig.~\ref{fig:MassDens} shows the mass-density diagram for the static
configuration of EOS~I and for three Kepler configurations with
different relative rotation-rates. The configurations to the right of the maximum are
on the so-called ``unstable branch'', because they will be subject to
unstable modes under small perturbations.
The configurations above the dotted line correspond to stars on the
unstable-branch of the static curve. They have no stable non-rotating
counter-part, even if they are on the stable branch of the mass-curve
of the rotating case, and they are therefore called ``supramassive
stars''. These configurations are stabilized by rotation and would
become unstable if slowed down below a critical rotation rate. 

\begin{figure}[htbp]
  \centering
  \includegraphics[width=0.4\textwidth,clip,angle=-90]{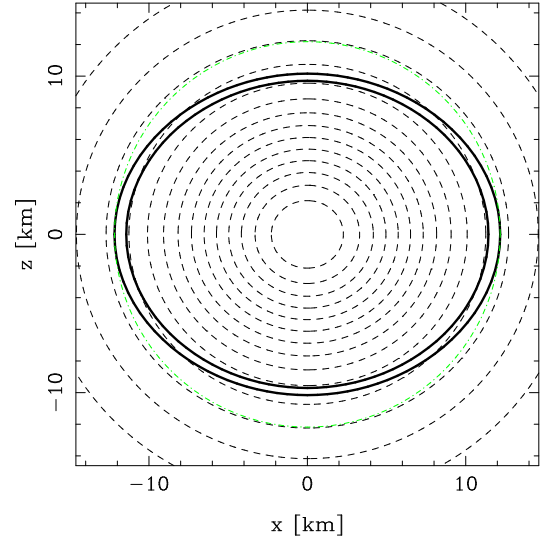}
  \caption{Configuration with protons rotating at the speed of the
  fastest known millisecond pulsar, $\Om_\p/2\pi=641$~Hz, and 
  $\Om_\n/2\pi=645$~Hz. The protons are extending further outside than
  the neutrons.
  The physical parameters are $\mu_\n=0.228\;m_b c^2$ and
  $\mu_\p=0.220\;m_b c^2$, resulting in central baryon number density
  $n_c=0.561\;\fmq$, proton fraction $\xp=0.09$ and a
  gravitational mass of $\Mg=1.39\;\Msol$.}
  \label{fig:msPulsar}
\end{figure}

So far we have considered stars chemical equilibrium at the center,
i.e. $\mu_\n = \mu_\p$.  
Incidentally, for the EOS-class considered here, the resulting static
configurations share a common outer surface in this case.
However, global chemical equilibrium is generally not possible for
configurations with the two fluids rotating at different rates, which
was shown by \citet{andersson01:_slowl_rotat_GR_superfl_NS} and
\citet{prix02:_slow_rot_ns_entrain}.  In order to model more ``realistic''
configurations, in which the proton-fluid  mimics a neutron-star
``crust'' (albeit without any solidity) by extending further
outside than the neutrons, we can easily achieve this by choosing
different central chemical potentials.
For example, using EOS~II and setting $\mu_\n = 0.228\;m_b c^2$ and
$\mu_\p=0.220\;m_b c^2$, we obtain the configuration shown in
Fig.~\ref{fig:msPulsar}. For this figure we have chosen the rotation
rate of the fastest known millisecond pulsar, which has a period of
$P\sim1.56$~ms. 

\begin{figure}[htbp]
  \centering
  \includegraphics[width=0.4\textwidth,clip,angle=-90]{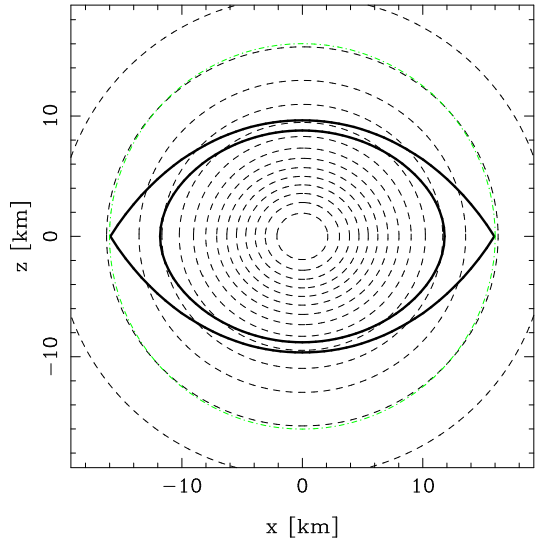}
  \caption{Kepler configuration with $\R=0.01$, $\mu_\n=0.28\;m_b c^2$ and
  $\mu_\p=0.3\;m_b c^2$, corresponding to a central baryon number density
  of $n_c=0.716\;\fmq$, proton fraction $\xp=0.09$ and a gravitational
  mass of $\Mg=1.57\;\Msol$. The protons extend to the outer
  surface. The maximal rotation rates are found as
  $\Om_\n/2\pi=924.5$~Hz and $\Om_\p/2\pi=915.3$~Hz.} 
  \label{fig:Kepler_II}
\end{figure}

A similar configuration with $\mu_\n=0.28\;m_b c^2$ and
$\mu_\p=0.3\;m_b c^2$ rotating at the Kepler-limit for a 
relative rotation rate of $\R=0.01$ is displayed in
Fig.~\ref{fig:Kepler_II}. 
As can be seen by the  cusp-formation, the Kepler-limit is determined
by the \emph{outer} fluid, i.e. the protons in this case, despite the
fact that they are rotating \emph{more slowly} than the neutrons. This 
contrasts with the case depicted in Fig.~\ref{fig:Kepler}, in which the
faster fluid also happens to be the outer fluid, which is a
particularity of this EOS-class and the choice of central chemical
equilibrium $\mu_\n=\mu_\p$ (cf. \cite{prix02:_slow_rot_ns_entrain}).

\section{Conclusions}
\label{sec:summary}

We have developed a theoretical framework and a numerical code for
computing stationary, fully relativistic superfluid neutron star models.

Using this code we have reconfirmed the existence of oblate-prolate
shaped two-fluid configurations, previously shown in
\cite{prix02:_slow_rot_ns_entrain,prix02:_relativistic_sf_ns}.
We have studied the dependency of the Kepler rate of a two-fluid star
on the relative rotation rate $\R$. We have compared this to the
Kepler-rate predicted by a Newtonian slow-rotation approximation,
which is found to typically overestimate the Kepler-rate by about
$15\%$, but which can also sometimes underestimate it, as seen in
the case of EOS~I and $\R\lesssim -0.25$ (cf. Fig.~\ref{fig:Kepler}).

The relative rotation rate can also have a large influence (at fixed
central density) on the mass-density relation, as shown in
Fig.~\ref{fig:MassDens}.

Another interesting aspect of this model is that we are not restricted
to configurations in chemical equilibrium at the
center. Choosing the central chemical potentials to be different
allows one to emulate a neutron star ``crust'' (albeit a fluid one),
as one fluid will now extend further outwards than the other, as seen
in Fig.~\ref{fig:msPulsar} and Fig.~\ref{fig:Kepler_II}. 
One interesting observation from such configurations is that the
Kepler-limit will be determined by the outer fluid (forming a cusp),
while this can actually be rotating \emph{slower} then the inner fluid.

We currently use a (quite general) EOS class of two-fluid polytropes,
but this can be extended straightforwardly to more ``realistic''
nuclear-physics equations of state. In particular it might be
interesting in the next step to use the first relativistic two-fluid
EOS incorporating entrainment by \citet{comer03:_relat_mean_field}.
Furthermore it would be important to add the presence of a solid crust
and to allow for differential rotation in the superfluid neutrons 
(differential rotation in single-fluid stars has been implemented and
used in LORENE already, cf. \cite{villain04:_evolut,goussard98:_rapid}). 

The astrophysically most interesting future extension of this work would
probably consist in studying the oscillation modes of such models,
which would be directly related to the emission of gravitational waves.
In these non-stationary situations, however, dissipative mechanisms
like viscosity and mutual friction would also start to play a role and
should be included in the model. 

\appendix

\section{The Newtonian analytic slow-rotation solution}
\label{sec:newt-analyt-slow}
 
A method for solving the stationary 2-fluid configuration in the
Newtonian slow-rotation approximation was initially developed in
\cite{prix99:_slowl_rotat_two_fl_ns}, and was completed to include all
EOS-interactions in \cite{prix02:_slow_rot_ns_entrain} (in the following referred to as
Paper~I). 
Using this method, an analytic solution was found in Paper~I for
equations of state of the form
 $\E = {1\over 2} \kapn n_\n^2 + {1\over 2} \kapp n_\p^2 + \kapnp n_\n n_\p + \beta^\p n_\p \Delta^2$. 
While this solution was very useful for studying the qualitative
properties of an interacting 2-fluid system, it is unfortunately not
very suitable for comparison to the numerical solution presented in
this paper. The reason for this lies in the somewhat unphysical
behavior of entrainment in this model.  
Namely, the entrainment numbers~(\ref{eq:10}) are found as 
$\eps_\p = {2 \beta^\p / m}$, and  $\eps_\n = 2 \xp \beta^\p / (m (1-\xp))$,
where $\xp\equiv n_\p/n$ is the proton fraction, which is constant for
this EOS (cf. Paper~I).
Therefore the entrainment numbers are constant, independently of the
densities, and so the entrainment effect would still be present in a 
1-fluid region.  This unphysical behavior does not pose a problem in
the slow-rotation approximation, which consists of an expansion around
a static chemical-equilibrium configuration: the two fluids share a common
surface in the unperturbed state, and the rotation will only induce
infinitesimal displacements of the fluids. In this framework there are
therefore no finite 1-fluid regions. However, in a numerical code
allowing for arbitrary rotations and deviations from chemical
equilibrium, such an entrainment model would be problematic.
The EOS-class~(\ref{eq:anEOS}) used in this work is therefore
preferable on both physical and numerical grounds.  

Fortunately, an analytic
solution can also be found for this physically preferable
EOS using the slow-rotation approach developed in  Paper~I.
This solution is very valuable for quantitative
comparisons with our numerical results presented in
section~\ref{sec:comp-newt-slow}.  Here we derive this new
analytic solution, skipping some of the more technical steps, which
have been explained in more detail already in Paper~I. 

Because of axisymmetry, the rotating solution only depends on the
spherical coordinates $r$ and $\theta$, while the static configuration
is assumed to be spherically symmetric.
The 2-fluid slow-rotation approximation proceeds by expanding any local
stellar quantity $Q$ as follows:
\begin{equation}
Q(r,\theta; \Om_\X) = Q^\stat(r) + \Om_\X \, Q^{\X\Y}(r,\theta)\,\Om_\Y + \O(\Om^4)\,,
\end{equation}
where here and in the following we automatically sum over
repeated constituent indices $\X=\n,\p$. 
We can separate the variables $r$ and $\theta$ by expanding in
Legendre Polynomials, i.e. $Q^{\X\Y}(r,\theta) = \sum_l Q^{\X\Y}_l(r) P_l(\cos\theta)$, 
and it can be shown that only the components $l=0, 2$ will be nonzero
in the solution. The solution is therefore fully determined by two
ordinary differential equations for the components $\Phi^{\X\Y}_0(r)$
and $\Phi^{\X\Y}_2(r)$ of the perturbation of the gravitational
potential.
The information about the EOS enters via the following two ``structure
functions'', defined as
\begin{equation}
\S_{\X\Y} \equiv \left(\left. {\partial^2 \E \over \partial n_\X \partial n_\Y} \right|_0\right)^{-1}\,,
\quad \beta^\X  \equiv \left. {\partial^2 \E \over \partial n_\X \partial \Delta^2} \right|_0 \,,
\end{equation}
where $|_0$ denotes the derivatives to be evaluated at the static
configuration. For the EOS (\ref{eq:anEOS}), we find 
\begin{equation}
\S_{\X\Y} = {1\over \K}\left(\begin{array}{c c}
\kapp & -\kapnp \\
-\kapnp &  \kapn \\
\end{array} \right)\,,
\end{equation}
where $\K \equiv \kapp \kapn - \kapnp^2$, and 
\begin{equation}
\beta^\X(r) = \kapD M^{\X\Y} n_\Y^\stat(r)\,,
\end{equation}
with the constant matrix $M^{\X\Y}$ defined as
\begin{equation}
M^{\X\Y}\equiv \left(\begin{array}{c c}
0 & 1 \\
1 & 0 \\
\end{array}  \right)\,.
\end{equation}
We further introduce the ``derived'' structure functions, 
\begin{equation}
k_\A \equiv \S_{\A\B}m^\B\,,\qaq k  \equiv m^\A k_\A\,,
\end{equation}
which are constant for this EOS. The matrices $E_\A^{\X\Y}$, defined as 
\begin{equation}
    E_\A^{\X\Y}(r) \equiv {1 \over 3} \S_{\A\B} \left(
    \delta^{\B,\,\X\Y} - 2 \beta^\B(r)\, \Delta^{\X\Y}\right) \,,
 \label{eq:EAXY}
\end{equation}
are now functions of $r$, contrary to the EOS treated in Paper~I, 
in which they were constant. The constant auxiliary matrices
$\delta^{\A,\X\Y}$ and $\Delta^{\X\Y}$ are defined as
\begin{eqnarray}
   \delta^{\n,\,\X\Y} &\equiv& \left(\begin{array}{c c} 
       m^\n & 0 \\
       0    & 0 
\end{array}\right) \,, \quad
   \delta^{\p,\,\X\Y} \equiv \left(\begin{array}{c c}
       0  & 0 \\
       0  & m^\p 
\end{array}\right) \,,\\
   \Delta^{\X\Y} &\equiv& \left(\begin{array}{rr} 
         1 & -1 \\
       - 1 &  1 \\
\end{array} \right) \,. \label{equConstDelta}
\end{eqnarray}
The \emph{static} background solution only depends on $\S_{\X\Y}$, and
is identical to the one found in Paper~I. Namely, in ``natural units''
defined by \mb{$\rho^\stat(0)=1$} and \mb{$R=1$}, this static solution
can be written as 
\begin{equation}
\rho^\stat(r) = {\sin(r\sqrt{k})\over r\sqrt{k} } \,. \label{eq:StaticSol}
\end{equation}
In these units it must be true that \mb{$\rho^\stat(1)=0$}, which leads
to the condition \mb{$k = m^\A k_\A = \pi^2$}. This relation can be
used to rescale the EOS parameters $\kapn$, $\kapp$ and $\kapnp$ to
natural units.
The respective particle number densities $n_\A^\stat(r)$ are expressible as
\begin{equation}
n_\A^\stat (r) = {k_\A \over \pi^2} \rho^\stat(r)\,. \label{eq:nAstat}
\end{equation}
Substituting this into (\ref{eq:EAXY}), we can write
\begin{equation}
E_\A^{\X\Y}(r) = \Et_\A^{\X\Y} - \Eh_\A^{\X\Y} \rho^\stat(r)\,,
\end{equation}
in terms of the two constant matrices
\begin{eqnarray}
\Et_\A^{\X\Y} &\equiv& {1\over 3} \S_{\A\B} \delta^{\B,\X\Y}\,, \nonumber\\
\Eh_\A^{\X\Y} &\equiv& {2 \over 3 \rho^\stat} \, \S_{\A\B} \beta^\B \,\Delta^{\X\Y}\,, 
\label{eq:DefEAXY}
\end{eqnarray}
and for \mb{$E^{\X\Y} \equiv m^\A E_\A^{\X\Y}$}, we write in an analogous manner
\begin{equation}
E^{\X\Y} = \Et^{\X\Y} - \Eh^{\X\Y}\, \rho^\stat(r)\,,
\end{equation}
with
\begin{eqnarray}
\Et^{\X\Y} &=& {1\over 3} k_\B \delta^{\B,\X\Y}\,,\nonumber\\
\Eh^{\X\Y} &=& {4 \kapD \over 3 \pi^2}\, k_\n k_\p  \,\Delta^{\X\Y}\,.
\label{eq:DefEXY}
\end{eqnarray}
We can now write the differential equations determining the solution
for the given EOS, namely  
\begin{eqnarray}
\D_0 \Phi^{\X\Y}_0 + \pi^2\,\Phi^{\X\Y}_0 &=& \C^{\X\Y} + r^2\Et^{\X\Y} - r^2 \rho^\stat\Eh^{\X\Y}\!\!, \hspace{0.8cm}\label{equODE0} \\
\D_2 \Phi^{\X\Y}_2 + \pi^2 \,\Phi^{\X\Y}_2 &=& -r^2 \Et^{\X\Y} +  r^2\rho^\stat \Eh^{\X\Y}, \label{equODE2}
\end{eqnarray}
where the differential operator $\D_l$ is defined as
\begin{equation}
\D_l \equiv  {d^2 \over dr^2} + {2 \over r} {d \over dr} - {l (l + 1) \over r^2}\,.
\end{equation}
We note that the only difference of this EOS to the one studied in
Paper~I concerns the entrainment $\beta^\X(r)$. We can therefore
formally recover the results from Paper~I in the limit
$\kapD\rightarrow0$, which corresponds to $\Eh^{\X\Y} \rightarrow 0$
and $\Eh_\A^{\X\Y}\rightarrow 0$.
The constant matrix $\C^{\X\Y}$ is determined by the choice of
stellar sequence, e.g. either characterized by fixed central densities
(FCD) or fixed masses (FM). 
The solution to the above equations also determines the density
distribution of the two-fluid star, namely via the relations
\begin{eqnarray}
n_{\A,0}^{\X\Y}(r) &=& \S_{\A\B}\,\C^{\B,\X\Y} + r^2 E_\A^{\X\Y} - k_\A \Phi_0^{\X\Y}\,,\label{eq:nA0}\\
n_{\A,2}^{\X\Y}(r) &=& - r^2 E_\A^{\X\Y} - k_\A \Phi_2^{\X\Y}\,,
\label{eq:nA2}
\end{eqnarray}
where the constants $\C^{\A,\X\Y}$ are also determined by the choice of
stellar sequence, and they satisfy the relation $\C^{\X\Y} = k_\A \C^{\A,\X\Y}$.
The complete slow-rotation solution for the density distribution of the
two fluids can be written as
\begin{equation}
n_\A(r,\theta) = n_\A^\stat(r) + \Om_\X\left( n_{\A,0}^{\X\Y} + n_{\A,2}^{\X\Y} P_2(\cos\theta)\right) \Om_Y\,.
\end{equation}
The general (regular) solution of equations (\ref{equODE0}) and
(\ref{equODE2}) can be found explicitly as
\begin{eqnarray}
\Phi_0^{\X\Y}(r) &=& \mathcal{A}^{\X\Y}_0\, {J_{1/2}(r\pi) \over \sqrt{r}} 
+ {\Et^{\X\Y}\over \pi^2}\left( r^2 - {6\over \pi^2}\right) + {\C^{\X\Y}\over \pi^2}\nonumber \\  
&& \hspace*{-1cm} - {\Eh^{\X\Y} \over 12\pi^4}\left\{3 r\pi \sin r\pi + (3-2\pi^2 r^2)\cos r\pi\right\}\,,
\label{eq:Phi0gen}\\
\Phi_2^{\X\Y}(r) &=& \mathcal{A}^{\X\Y}_2\,{J_{5/2}(r \pi) \over
  \sqrt{r}} - {\Et^{\X\Y}\over \pi^2} r^2 \nonumber\\
&& - {\Eh^{\X\Y} \over 12 \pi^7 r^3}\left\{ (45 + 2\pi^4 r^4)\, r\pi \,\cos r\pi \right.\nonumber\\
&& \hspace*{2cm}\left. + 15 (r^2\pi^2 - 3) \, \sin r\pi \right\}\,,
\label{eq:Phi2gen}
\end{eqnarray}
where $\mathcal{A}_0^{\X\Y}$ and $\mathcal{A}_2^{\X\Y}$ are constants
of integration, and $J_n(x)$ are the standard Bessel functions. 
One can verify the asymptotic behavior \mb{$\Phi^{\X\Y}_2\sim r^2$} as
\mb{$r\rightarrow0$}, which is required for regularity. In addition to
the regularity requirements at the center, the solution must satisfy
the following boundary condition at the surface ($r=1$):
\begin{equation}
\Phi_l^{\X\Y'}(1)+l (l+1)\Phi_l^{\X\Y}(1)=0\,.
\end{equation}
These boundary conditions result in the following relations for the
integration constants $A^{\X\Y}_0$ and $A^{\X\Y}_2$:
\begin{eqnarray}
\hspace*{-0.5cm}4\pi^4 \sqrt{2} \, \mathcal{A}^{\X\Y}_0 &=& 12 (\pi^2 - 2) \Et^{\X\Y} + (1-\pi^2) \Eh^{\X\Y} \nonumber\\
&& + 4\pi^2 \C^{\X\Y}\,,\\
\sqrt{2}\, \mathcal{A}^{\X\Y}_2 &=& {5 \over \pi^2} \Et^{\X\Y} - {5 \over 12 \pi^4} ( 3 + 2\pi^2) \Eh^{\X\Y}\!\!.
\end{eqnarray}

\subsubsection*{Fixed central density (FCD) sequence}

The FCD-sequence is the most directly comparable to the numerical
results discussed in this paper. This sequence is defined by the
condition $n_{\A,0}^{\X\Y}(0)=0$, and in this case the remaining
constant of integration can be determined as 
\begin{eqnarray}
\C_\FCD^{\X\Y} &=& -3 \left(1 - {4\over \pi^2} \right)\, \Et^{\X\Y} + {1\over 4} \Eh^{\X\Y}\,,
\end{eqnarray}
and we also have the relation
\begin{equation}
  \label{eq:9}
  \S_{\A\B} \C_\FCD^{\B,\X\Y} = {k_\A \over \pi^2}\, \C_\FCD^{\X\Y}\,.
\end{equation}
Putting all the pieces together, we arrive at the following explicit
solution for the density perturbations of the FCD sequence:
\begin{eqnarray}
n_{\A,0,\FCD}^{\X\Y} &=& - {6 k_\A \Et^{\X\Y} \over \pi^4} \left(
  {\sin{r\pi} \over r\pi}  + {r^2\pi^2 \over 6} - 1\right) + \Et_\A^{\X\Y} r^2 \nonumber\\
&&\hspace*{-1.5cm}- {k_\A \Eh^{\X\Y} \over 4 \pi^4} \left( (1-r^2\pi^2) {\sin{r\pi}\over r\pi} 
  - (1- {2\over3} r^2\pi^2) \cos{r\pi} \right) \nonumber\\
& & - {\Eh_\A^{\X\Y} \over \pi^2} r\pi \sin{r\pi}\,,
\end{eqnarray}
\begin{eqnarray}
n_{\A,2}^{\X\Y} (r) &=& {k_\A \Et^{\X\Y} \over \pi^2} \left(r^2 - {5\over \sqrt{2}}
{J_{5/2}(r\pi) \over \sqrt{r}} \right) - \Et_\A^{\X\Y}\, r^2\nonumber\\
&& \hspace*{-1.5cm} + {5 \over 6} {k_\A \Eh^{\X\Y} \over \pi^5 r^3}
\biggl\{ \left({\pi^2\over 5} r^4 - 3\right)r\pi\cos r\pi - (r^2\pi^2 - 3) \sin r\pi \biggl\}
\nonumber\\
&& + {\Eh_\A^{\X\Y} \over \pi^2} \,r\pi \sin r\pi\,,
\end{eqnarray}
in terms of the constant ``structure matrices'' $\Et$ and $\Eh$
defined in Eqs.~(\ref{eq:DefEAXY}) and (\ref{eq:DefEXY}).

\subsubsection*{Fixed-mass (FM) stellar sequence}

For completeness we also give the solution corresponding to a
fixed-mass sequence, which might even be physically more
interesting. The difference to the FCD-solution only concerns the
$l=0$ component, while $n_{\A,2}^{\X\Y}$ is the same in both cases. As
discussed in Paper~I, the FM-sequence is characterized by the
conditions 
\begin{equation}
  \label{eq:77}
  \int_0^1 r^2\, n_{\A,0,\FM}^{\X\Y}(r)\, d r = 0\,,
\end{equation}
which lead to the following condition for the potential
\begin{equation}
  \Phi_{0,\FM}^{\X\Y} (1) = 0\,.
\end{equation}
This results in the integration constant
\begin{eqnarray}
  \C^{\X\Y}_\FM = \left( {6\over\pi^2} - 1 \right)\, \Et^{\X\Y} +
  \left( {1\over6} - {1\over 4 \pi^2}\right)\, \Et^{\X\Y}\,,
\end{eqnarray}
while we can similarly determine $\S_{\A\B}\C^{\B,\X\Y}$ from (\ref{eq:77}).
Inserting this into (\ref{eq:Phi0gen}) we get
\begin{eqnarray}
  \Phi^{\X\Y}_{0,\FM} (r) &=& {\Et^{\X\Y} \over \pi^2} \left( r^2 -1 +
  \sqrt{2} {J_{1/2}(r \pi) \over \sqrt{r}} \right) \nonumber\\
&&\hspace*{-1cm}+ {\Eh^{\X\Y} \over 12 \pi^4} \biggl( 2 \pi^2 -3 + (2\pi^2 r^2 -3)\,\cos r\pi \nonumber\\
  & & - (1 + 3 r^2) {\pi^2\over \sqrt{2}} { J_{1/2}(r\pi) \over \sqrt{r}} \biggl)\,.
\end{eqnarray}
The $l=0$ density coefficient is therefore found by using (\ref{eq:nA0}):
\begin{eqnarray}
n_{\A,0,\FM}^{\X\Y} &=& {k_\A \Et^{\X\Y} \over 5 \pi^4} 
\left(30 + 3 \pi^2 - 5 r^2\pi^2 - {10\pi \over r} \sin{r\pi} \right) \nonumber\\
&& \hspace*{-1cm}+ \Et_\A^{\X\Y} \left( r^2  - {3\over5} \right) 
+ {\Eh_\A^{\X\Y}\over \pi^2} \left(3(1- {6\over \pi^2}) - r\pi \sin r\pi \right )\nonumber\\
&& \hspace*{-0.5cm} -{k_\A \Eh^{\X\Y} \over 12 \pi^4}
\left( 36 (1 - {6\over \pi^2}) - (3 - 2r^2\pi^2) \cos{r\pi} \right.\nonumber\\
&&\left. - (1 + 3 r^2) {\pi\over r}\sin{r\pi} \right) \,,
\end{eqnarray}
which completes the analytic solution in the FM case.

\subsubsection*{Calculating the Kepler-limit}

We briefly review the method of calculating the Kepler-limit using the
slow-rotation solution presented above. The result of this calculation
was used for the Newtonian slow-rotation Kepler-limit presented in
Fig.~\ref{fig:Kepler}. 
As derived in Paper~I, the Kepler-rate to order $\Om^2$ for each of
the two fluids can be expressed as the solution of the equation 
\begin{equation}
  \label{eq:11}
  \Om_\A^2 = \Om_\stat^2 + \Om_\X \,\d q_\A^{\X\Y} \,\Om_\Y + \O(\Om^4)\,,
\end{equation}
where the zeroth-order expression is
\begin{equation}
  \label{eq:13}
  \Om_\stat^2 = \Phi^{\stat'}(1)\,,
\end{equation}
and the second-order correction terms reads as
\begin{equation}
  \label{eq:12}
\d q_\A^{\X\Y} = \left[ - {3 \over k_\A} \left(n_{\A,0}^{\X\Y} - {1\over2} n_{\A,2}^{\X\Y}
 \right) + \Phi_0^{\X\Y'} -{1\over2} \Phi_2^{\X\Y'} \right]_{r=1}\,.
\end{equation}
For the EOS-class considered here, we find
\begin{eqnarray}
  \label{eq:14}
  \Om_\stat^2 &=& {4\over \pi}G \, \rho(0)\,, \\
  \d q^{\X\Y}_{\A,\FCD} &=& - {9 \Et_\A^{\X\Y} \over 2 k_\A} +
  {\Eh^{\X\Y}\over 12 \pi^4} \left( 6 - 7 \pi^2 \right) \nonumber\\
&& \hspace{0.5cm}+ {\Eh^{\X\Y}\over \pi^4}\left( 5 \pi^2 - 24 \right)\,,\\
\d q^{\X\Y}_{\A,\FM} &=& {6 \over \pi^2} \left( {1\over 5} - {3\over
  \pi^2}\right) \Et^{\X\Y} - {9 \Eh_\A^{\X\Y} \over k_\A \pi^4} (\pi^2 - 6)\nonumber\\
&& \hspace*{-1.0cm}- {\Eh^{\X\Y} \over 4 \pi^6} \left(216 - 39 \pi^2 + 2 \pi^4\right) 
- {27 \over10 k_\A} \Et_\A^{\X\Y} \,.
\end{eqnarray}
For each fluid $\A$, we find the Kepler-limit
$\Om_{\Kepler,\A}(\Om_\B)$ (where $\B\not=\A$) by solving the quadratic
equation (\ref{eq:11}). The Kepler-limit is then interpreted as
the corresponding solution for the \emph{faster} fluid, which in this
case corresponds to the outer fluid, i.e.  we have
\begin{equation}
  \label{eq:15}
  \Om_{\Kepler} = \left\{\begin{array}{c c}
      \Om_{\Kepler,\n}(\R)\,,\quad\textrm{for}\quad \R > 0\,,\\
      \Om_{\Kepler,\p}(\R)\,,\quad\textrm{for}\quad \R < 0\,.
    \end{array}\right.
\end{equation}

\begin{acknowledgments}
RP acknowledges support from the EU Programme 'Improving the Human
Research Potential and the Socio-Economic Knowledge Base' (Research
Training Network Contract HPRN-CT-2000-00137).  GC acknowledges 
support from NSF grant PHYS-0140138.
\end{acknowledgments}

\bibliography{biblio}

\end{document}